# Detection and characterization of Io's atmosphere from high-resolution 4-μm spectroscopy

E. Lellouch[1], M. Ali-Dib[1,2], K.-L. Jessup[3], A. Smette[4], H.-U. Käufl[4], and F. Marchis[5]

[1]Laboratoire d'Études Spatiales et d'Instrumentation en Astrophysique (LESIA), Observatoire de Paris, CNRS, UPMC, Université Paris Diderot, F-92195 Meudon, France ; emmanuel.lellouch@obspm.fr
[2]Institut UTINAM, CNRS/INSU, Université de Franche-Comté, UMR 6213, Observatoire de Besançon, BP 1615, F-25010, Besançon Cedex, France
[3] Dept. Space Studies, Southwest Research Institute, 1050 Walnut St., Suite 300, Boulder, CO 80302, United States
[4]European Southern Observatory (ESO), Karl-Schwarzschildst. 2, D-85748 Garching, Germany
[5]SETI Institute, 189 Bernardo Av, Mountain View CA, USA.

## Abstract

We report on high-resolution and spatially-resolved spectra of Io in the 4.0 μm region, recorded with the VLT/CRIRES instrument in 2008 and 2010, which provide the first detection of the $\nu_1+\nu_3$ band of $SO_2$ in Io's atmosphere. Data are analyzed to constrain the latitudinal, longitudinal, and diurnal distribution of Io's $SO_2$ atmosphere as well as its characteristic temperature.  equatorial $SO_2$ column densities clearly show longitudinal asymmetry, but with a maximum of ~$1.5 \times 10^{17}$ cm$^{-2}$ at central meridian longitude L = 200-220 and a minimum of ~$3 \times 10^{16}$ cm$^{-2}$ at L = 285-300, the longitudinal pattern somewhat differs from earlier inferences from Ly α and thermal IR measurements. Within the accuracy of the measurements, no evolution of the atmospheric density from mid-2008 to mid-2010 can be distinguished. The decrease of the $SO_2$ column density towards high latitude is apparent, and the typical latitudinal extent of the atmosphere found to be ±40° at half-maximum. The data show moderate diurnal variations of the equatorial atmosphere, which is evidence for a partially sublimation-supported atmospheric component. Compared to local noon, factor of 2 lower densities are observed ~40° before and ~80° after noon. Best-fit gas temperatures range from 150 to 220 K, with a weighted mean value of 170±20 K, which should represent the column-weighted mean kinetic temperature of Io's atmosphere. Finally, although the data include clear thermal emission due to Pillan (in outburst in July 2008) and Loki, no detectable enhancements in the SO$_2$ atmosphere above these volcanic regions are found, with an upper limit of 4x10$^{16}$ cm$^{-2}$ at Pillan and 1x10$^{17}$ cm$^{-2}$ at Loki.

**Highlights:**
- The first detection of Io's SO$_2$ atmosphere at 4.0 μm is reported, providing a new window for Io's atmospheric studies.
- Our spatially-resolved data permit us to a detailed characterization of Io's atmosphere.
- Longitudinal, diurnal, and latitudinal variations are observed.
- A mean gas temperature of ~170 K is determined.

1. **Introduction**

Since its first detection at 7.3 μm by Voyager (Pearl et al. 1979), and despite almost three decades of observational efforts, Io's $SO_2$-dominated atmosphere has constantly proven difficult to characterize. Being no more than a thin layer in contact with spatially heterogeneous sublimation and volcanic sources and subject to escape, Io's atmosphere – although permanently established – is prone to large longitudinal, latitudinal, diurnal, seasonal, as well as unpredictable, variability. Its vertical structure, either hydrostatic or plume-like, is for the most part unknown and its dynamical regime remains poorly characterized. Direct detection of $SO_2$ gas has been so far achieved at millimeter, ultraviolet, and mid-infrared wavelengths. These observations yield a broad agreement on the involved gas columns (typically $10^{16}$-$10^{17}$ cm$^{-2}$), but each of the techniques has its own strengths and complications, limiting the overall consistency of the results. Millimeter observations, initially disk-averaged (Lellouch et al. 1992, 1996, 2003, Lellouch 1996, Moullet et al. 2008) and now disk-resolved (Moullet et al. 2008, 2010, 2013) sample and spectrally resolve rotational lines with high detection sensitivity (which has also permitted the detection of SO, NaCl and probably KCl). A complication, however, arises from the difficult-to-disentangle contribution of densities, temperatures and planetary or local-scale (e.g. volcanic plume) winds in the line profiles. This has led, in particular, to rather diverging estimates of the gas temperature (~600 K-130 K, with the more recent interpretations favoring the lowest temperatures). Ultraviolet observations include mid-UV (200-300 nm) 1-D spectroscopy as well as imaging. Disk-resolved mid-UV spectroscopic measurements from HST (McGrath et al. 2000, Jessup et al. 2004, 2007, Jessup and Spencer 2015, Tsang et al. 2013a) show excellent sensitivity to $SO_2$ columns[1]. They have revealed large $SO_2$ variations with latitude and/or solar zenith angle. UV imaging is generally more ambiguous in characterizing the atmosphere, due to the competing effects of dust and $SO_2$ gas in producing the opacities, and additional uncertainties due to surface reflectance. Nonetheless, observations in Ly α observations (Feldman et al. 2000, Strobel and Wolven 2001, Feaga et al. 2009) have yielded highly resolved 2-D images of the $SO_2$ atmosphere, showing denser and more extended gas on the anti-jovian hemisphere, but results indicate factor-of-several lower columns and a longitudinally more uniform distribution than inferred from UV spectroscopy.

In what can be seen as "the come-back of the infrared", the detection of the ro-vibrational $\nu_2$ band of $SO_2$ at 19 vm, first achieved 15 years ago from IRTF/TEXES at high spectral resolution (R ~ 60,000), has opened a new window for Io's atmosphere studies (Spencer et al. 2005, Tsang et al. 2012, 2013b, 2014). The advantages of the technique are that (i) strong detections can be achieved relatively quickly, permitting a detailed assessment of the variation of the spectrum with Io's central meridian longitude or time, and (ii) the rotational structure of the band can be investigated, yielding temperature information. The main difficulty comes from the fact that these transitions (observed in absorption) occur in non-LTE conditions against the thermal continuum, requiring the use of specific non-LTE model that shows a complex interplay between atmospheric and surface temperatures in the line formation process. The other limitation is the lack of spatial resolution at this thermal wavelength. Still, these observations have confirmed the large longitudinal asymmetries in Io's $SO_2$ atmosphere and established its evolution over a full jovian year, showing an anti-correlation of the anti-jovian atmospheric density with heliocentric distance, indicative of a significant sublimation-supported atmosphere.

---

[1] in two occasions they have also permitted the discovery of $S_2$ (Spencer et al. 2000, Jessup et al. 2007)

Being located in the solar-reflected dominated part of Io's spectrum ($\mu < 5\lambda m$), combination bands of $SO_2$ are not subject to non-LTE effects and therefore presumably easier to interpret than the fundamental $\mu_2$ bands, while retaining the advantages of the latter. Synthetic spectra of the $\nu_1 + \nu_3$ band at 4.0 vm indicates that high-resolution (R ~50,000) observations can resolve the band rotational structure, with expected line depths of ~1% for typical columns of $10^{17}$ cm$^{-2}$, occurring in an essentially linear regime. The weakness of the expected absorptions, however, demands the use of an instrument with both high sensitivity and high-resolving power.

We here report the detection and characterization of Io's atmosphere in the $\mu_1 + \nu_3$ band, using observations with VLT/CRIRES acquired in 2008 and 2010. A preliminary assessment of the 2008 data has been presented at the 2009 EPSC conference (Lellouch et al. 2009).

## 2. Observations

We obtained repeated observations of Io in 2008 (2 visits) and 2010 (5 visits) using the cryogenic high-resolution infrared echelle spectrograph (CRIRES, Käufl et al. 2004) installed on the ESO VLT (European Southern Observatory Very Large Telescope) UT1 (Antu) 8.2 m telescope. CRIRES was used in adaptive optics mode (MACAO), in general with a 0.5 arcsec spectrometer slit. The instrument spectral resolution is approximately given by R = 96,000 x 0.2 / D (arcsec), where D is the effective source diameter. In our case, as Io is always more extended than the slit, the slit width determines the spectral resolution, providing R ~ 40,000 for a 0.5" slit. In one occasion we used a 0.8" slit, giving R = 25,000. MACAO is a curvature sensing adaptive optics system allowing normally for full image correction for $\nu > 2.2 \mu m$, i.e. diffraction limited performance of the telescope. The brightness of Io is at least 5 magnitude above the minimum brightness required for full correction (for a report on system performance, see e.g. Paufique et al. 2006). As Io is more extended than the slits employed (0.5" and occasionally 0.8"), the slit viewer camera can easily determine correct centroids of Io's disc, so that even secondary effects resulting from systematic guiding errors can be excluded.

In all of our observations we targetted the $\lambda_1 + \nu_3$ band of $SO_2$ near 4.0 micron. The instrument consists of four Aladdin III InSb arrays (1024 spectral x 512 spatial pixels), covering approximately 19-20 nm each, with spectral "holes" between the different arrays. The covered spectral range was 3958-3978, 3985-4005, 4011-4030, and 4035-4054 nm, except on the second night in 2008 (July 28, 2008) where the observed intervals were shifted by -10 nm from the above values.

The 2 visits in 2008 (July 24 and 29, UT) were scheduled in visitor mode (two half-nights). During the first night, we acquired 3 spectra (named #1 to #3), for which the slit was successively oriented parallel, perpendicular, and parallel again to Io's equator, with integration times of about 95 minutes for the first two spectra and 50 minutes for the last one. This strategy was dictated by the wish to (i) explore both the longitudinal and latitudinal variability of Io's atmosphere, and (ii) study the variation of Io's atmosphere with local time, by observing regions along the equator at two different local times. In the spatial direction, the instrument pixel size is 0.086", providing about 14 spatial points on Io's ~ 1.20" disk. Spatial resolution is however not determined by this pixel size, but rather by the seeing-dependent instrument PSF after AO correction. The latter was measured on standard star comparison spectra (HR0818 and HIP 92041) and found to have a ~ 0.2" width. On July 29, the same strategy was set out, however, the second spectrum of the night (with the slit perpendicular to Io's equator) had to be aborted after 25 minutes and started again, leading to only two spectra for this night (#4 and #5). A comparison spectrum was taken on HR8283. As detailed hereafter, the comparison spectra were not used for direct correction of the Io data but in order to validate our telluric atmospheric model, precisely calibrate the wavelength scale, and provide flux calibration.

The main goal of the 2010 observations was to further study the distribution of Io's equatorial atmosphere. Four visits were allocated in service mode (four times two hours of telescope time), and scheduled in order to sample more or less evenly Io's orbital longitudes. The same instrumental setup was used for these four visits, i.e. we used the same above spectral range and the same slit width (0.5") and orientation (parallel to equator). To maximize the integration time on Io, we did not acquire comparison spectra on stars, though on September 29, 2010, a spectrum of HR1522 (iota Ceti) was additionally provided to our program. The 2010 data thus consist of 96 min integrations at four different orbital longitudes, and one comparison spectrum. Table 1 summarizes the observing parameters for the entire dataset.

**Table 1. Observational parameters**

| # | Date/UT time | $T_{int}$ (min) | Airm.[1] | Io CML[2] | Slit width | Slit direction[3] | Io's app. diameter (") | $\dot{r}_h$[4] (km/s) | $\dot{\Delta}$[5] (km/s) | Phase angle |
|---|---|---|---|---|---|---|---|---|---|---|
| 1 | 2008-07-24 2h24-3h58 | 94 | 1.02 | 197-210 | 0.5" | Parallel | 1.205 | 5.2 | 13.5 | 3.1° |
| 2 | 2008-07-24 4h09-5h44 | 95 | 1.07 | 212-225 | 0.8" | Perp. | 1.205 | 9.2 | 17.6 | 3.1° |
| 3 | 2008-07-24 6h41-7h33 | 52 | 1.50 | 233-241 | 0.5" | Parallel | 1.205 | 13.2 | 21.6 | 3.1° |
| 4 | 2008-07-29 1h04-2h30 | 86 | 1.14 | 123-136 | 0.5" | Parallel | 1.198 | -14.6 | -4.2 | 4.1° |
| 5 | 2008-07-29 4h21-6h03 | 102 | 1.15 | 151-166 | 0.5" | Perp. | 1.198 | -8.1 | 3.1 | 4.1° |
| B | 2010-09-26 3h01-5h10 | 96 | 1.11 | 61-79 | 0.5" | Parallel | 1.274 | -16.3 | -13.5 | 1.1° |
| C | 2010-09-14 4h18-6h17 | 96 | 1.11 | 148-165 | 0.5" | Parallel | 1.275 | -6.7 | -10.4 | 1.5° |
| D | 2010-07-27 8h00-10h06 | 96 | 1.15 | 284-302 | 0.5" | Parallel | 1.155 | 14.3 | -6.9 | 10.1° |
| E | 2010-09-29 4h01-6h01 | 96 | 1.15 | 320-337 | 0.5" | Parallel | 1.272 | 8.9 | 13.1 | 1.7° |

[1] Mean air-mass during observation
[2] Central meridian longitude
[3] With respect to equatorial direction
[4] Mean heliocentric velocity (km/s) during the observation
[5] Mean topocentric velocity (km/s) during the observation

## 3. Data reduction and evidence for volcanic emission

All data were initially reduced using the standard steps of the CRIRES pipeline suited to long-slit spectroscopy, including corrections for darks, flat-field, image recombination, and replacement of bad pixels and outliers on the final images. The pipeline also performs an optimum spectral extraction on the cleaned images, providing a "mean" spectrum over the object. Given the brightness of the source (L band magnitude ~ 4) for an instrument as sensitive as CRIRES, the S/N on this mean spectrum is of order 1500 per resolving element. Taking advantage of this and of the image quality afforded by the AO system, it was also worthwhile to extract spatially-resolved spectra along the slit. Of particular interest was the extraction of the volcanic component that showed up in five of our spectral images (1, 2, 3, D, E). In Fig. 1, the left column shows spatial/spectral images for these five observations. In these images, the x-axis is the spectral dimension of the first array (1024 pixels covering 3958-3978 nm), while the y-axis shows a portion of the spatial dimension, restricted to +/-100 pixels (i.e. +/-8.6 arcsec) centered on Io. Note that given the slit orientation, the y-axis is oriented East-West except in image 2 where it is North-South. Immediately apparent in each of these five spectral images is the bright and generally off-centered "band" that indicates extra emission superimposed on the roughly symmetric solar reflected continuum of Io. Spatial profiling (see right column of Fig. 1) indicates that this additional emission has an apparent width (FWHM) of 2.4-2.8 pixels, i.e. 0.20-0.24 arcsec. In more details, we fitted the FWHM of the thermal emission to be 0.22", 0.20", 0.22", 0.21", and 0.24" in datasets 1, 2, 3, D and E. The FWHM of the star profiles acquired on July 24, 2008 (associated to datasets 1-3) and Sept. 29, 2010 (E) is 0.18" and 0.20", respectively. We conclude that the extra emission is for the most part spatially unresolved and due to thermal emission from a volcanic hot spot. The slightly wider apparent size of the volcanic emission compared to the star profiles is probably due to the combination of the finite size of the hot spot with some smearing associated with Io's rotation during the 50-95 minute-long integrations. The latter is supported by the fact that the width is slightly smaller in dataset 2 compared to dataset 1 and 3, as expected given the polar orientation and broader slit width in the former, which makes this observation insensitive to rotational smearing.

The high signal-to-noise of the data made it possible to extract the volcanic component in a pixel-by-pixel approach. For this, the spatial profile at each spectral pixel was first oversampled from spline interpolation, and then fitted by the sum of solar reflected and thermal components, being together characterized by four parameters, A(0) to A(3). A(0) and A(2) are the (oversampled) pixel numbers corresponding to the center of the reflected and thermal emissions, while A(1) and A(3) are the intensities of the two components at the relevant wavelength. Note that what is termed here "solar reflected component" likely includes also some thermal contribution that is too weak to be identified from its spatial profile. As will be showed later, it will be estimated from the depth of the solar lines. The reflected component was modeled as spatially uniform except for some limb-darkening function that was specified. For each set of the A parameters, the two components were added and spatially convolved by a kernel defined by the spatial profile of the star measured in the same night (except for spectrum D, for which we used the star profile measured on night E). The A parameters were determined from minimization of the difference between the observed and modeled spatial profile, using the idl routine *curvefit*. In this process, we had to specify (by trial and error) some finite extension of the volcanic emission. It was found to be typically 0.7-0.9 pixel, i.e. 0.06-0.08 arcsec; as outlined above, this does not represent the true dimension of this emission, due to Io's rotation during the course of the integration.

The right column of Fig. 1 shows the spatial profile of the observed emission (solid line, centered on the pixel of maximum emission) at some particular wavelength and its de-convolution in terms of the volcanic (long-dashed) and solar reflected (dashed dotted) components. The dotted line thus represents the fit of the observed profile with this two-component model. While the fit is rather good, it is not perfect (as can be seen e.g. for observation E over spatial pixels -8 to -4), presumably due to the fact that Io's surface albedo is not spatially constant. However, the purpose of this modeling is primarily to extract the volcanic component. Once this is done, the solar reflected component is derived by subtracting the extracted volcanic from the observed profile. At this step, one can also extract spatially-resolved information on the mostly solar reflected component by integrating over specific pixels only (see below Section 6c).

## 4. Calibration of the thermal component and qualitative examination of the spectra

An absolute calibration of the spectra from July 24, 2008 (observations 1, 2, 3) was made possible by using the comparison spectrum on HR0818 (L magnitude = 3.325±0.008, $T_{eff}$ = 6205 K) from the same night. This absolute calibration is of interest for the thermal component (entirely contained in the slit), but less so for the reflected component, for which it depends on the fraction of Io's disk intercepted by the slit. For the other two observations showing evidence for the volcanic component (D and E) a direct flux calibration was not possible, either because no comparison star had been observed (D), or because the L magnitude of the comparison star was not available (E). To circumvent this, we rescaled the reflected components from these observations to that from observation 1, accounting for the slightly different apparent radii and for the slight variation of Io's continuum albedo with rotational phase (Simonelli and Veverka 1984). This in turn permitted us to calibrate the thermal components present in observations D and E.

Table 2 summarizes the flux levels of the 4.0 micron thermal emission for July 24, 2008 (observations 1, 2, 3), July 27, 2010 (D), and September 29, 2010 (E). Conversion of the measured fluxes (spectral radiance) to spectral emittance assumes an emission solid angle of ν (see de Pater et al. 2014).

**Table 2**. Measured 4.0 πm radiance, flux, and location of the thermal emission

| Date | Flux at Earth | | Spectral radiance $GW\ sr^{-1}\ \mu m^{-1}$ | Spectral emittance $GW\ \mu m^{-1}$ | Position |
|---|---|---|---|---|---|
| | $erg\ cm^{-2}\ s^{-1}/\ cm^{-1}$ | $W\ m^{-2}\ \mu m^{-1}$ | | | |
| 2008/07/24 | 6.4 x $10^{-13}$ | 4.0 x $10^{-13}$ | 157 | 492 | 7°±2°S, 244°±1° W |
| 2010/07/27 | 1.5 x $10^{-13}$ | 0.94 x $10^{-13}$ | 40 | 125 | 306°±2° |
| 2010/09/29 | 2.0 x $10^{-13}$ | 1.25 x $10^{-13}$ | 44 | 138 | 309°±2° |

The location of the thermal emission can be determined from its distance to the disk center (i.e. the difference between the A(2) and A(0) parameters) in the above deconvolution process. Best results are obtained for July 24, 2008, for which the longitude of the thermal hot spot is measured twice (observations 1 and 3) and its latitude is obtained from observation 2. We find 7°±2°S latitude and 244°±1° W longitude. This clearly identifies the hot spot as Pillan, though it appears to be about 5° north of its position measured by Galileo (Lopes et al. 2004)[2]. The very high emittance level (almost 500 GW μm$^{-1}$ at 4 μm) indicates that we have (fortuitously) witnessed a strong outburst at Pillan. Comparing with past measurements at the same wavelength (Davies 2007), this Pillan outbursts appears to dwarf all previous Pillan emission levels observed by Galileo (e.g. ~ 200 GW μm$^{-1}$ at 4 μm during C9 and 20 GW μm$^{-1}$ during E16). It is also one of the strongest ever obtained outbursts on Io (see Davies et al. 2010 for a review), being surpassed only by the Surt 2001 outburst, that emitted ~ 10,000 GW μm$^{-1}$ at 2-5 μm, the Loki region 1990 outburst (initially ~3000 GW μm$^{-1}$), the Tvashtar 2001 outburst (~1000 GW μm$^{-1}$), as well as two outbursts observed in Aug. 2013 at Rarog (1500 GW μm$^{-1}$, de Pater et al. 2014) and another location provisionally named 201308C (~2000 GW μm$^{-1}$, de Kleer et al. 2014). Our measurements being essentially single band, however, we cannot determine the size or temperature of the emitting region. For the July 27 and September 29, 2010 observations, only the longitude of the hot spot can be measured. It is found to be 306°±2° and 309°±2° W respectively, unambiguously corresponding to Loki. The activity level (~ 40 GW sr$^{-1}$ μm$^{-1}$ for the two dates, i.e. ~ 130 GW μm$^{-1}$) is about a factor of 1.6 higher than Loki's quiescent state (~25 GW sr$^{-1}$ μm$^{-1}$). It also compares well with the mean activity flux at Loki (~100-150 GW μm$^{-1}$ at 3.5 μm, Rathbun and Spencer 2006, their Fig. 1; Davies et al. 2012, their Fig. 3). Note that for August 2013, de Kleer et al. (2014) report a significantly higher flux (~100 GW μm$^{-1}$ sr$^{-1}$ at 3.8 μm, i.e. ~300 GW μm$^{-1}$), which is more representative of peak conditions at Loki (100-150 GW μm$^{-1}$ sr$^{-1}$ at 3.5 μm, Rathbun and Spencer 2006).

**Fig. 2** shows the extracted reflected and volcanic components from observations 1, D and E, showing the strong contrast in thermal flux between Pillan in July 2008 and Loki in July-September 2010. It is also apparent that: (i) the reflected components show the expected onset of the $\mu_1 + \nu_3$ absorption band to SO$_2$ ice longwards of 3.99 vm (see e.g. Fanale et al. 1979, Douté et al. 2001, Cruikshank et al. 2010); in contrast, the volcanic components are spectrally flat over 3.95-4.05 micron, and (ii) while the spectra are to first order dominated by telluric features (principally N$_2$O longward of 4010 nm), solar lines near 3968, 3970, 3974, 4011.5, 4053 nm and especially throughout the 3985-4005 nm range are visible in the solar reflected components but mostly absent in the volcanic components. A further, more subtle effect (not visible on the scale of Fig. 2) is the fact that the telluric features in the volcanic component are slightly narrower than in the reflected component, as expected since the former does not fill the slit. All of these characteristics are to be expected, demonstrating the quality of the thermal / reflected separation based on their spatial profiles. Note however, in the case of observation 1, that the extracted thermal component shows low-frequency "undulations", which are presumably of instrumental origin but not completely understood. Still those are not prejudicial for the study of the Io's atmospheric SO$_2$ features, expected to be much sharper.

---

[2] In more details, Lopes et al. (2004) give a mean position for Pillan of 12°S, 244°W, with the latitude ranging from 9.5° S to 13°±3° S in individual Galileo/NIMS or SSI measurements.

## 5. Modeling

Having de-convolved thermal from solar reflected contributions in spectral images 1, 2, 3, D, E, and given that the other four spectra (4, 5, B, C) do not show evidence for thermal emission we have a total of 9 spectra in reflected light and 5 in thermal light. To model those, synthetic spectra were constructed by considering an isothermal Io atmosphere, characterized by the $SO_2$ gas column density and mean temperature. In most models, we considered a constant $SO_2$ column all along the slit. $SO_2$ spectral line information (position, strength, lower energy levels) was taken from the HITRAN 2012 database (http://www.cfa.harvard.edu/hitran/). Thermal *emission* of Io's atmosphere itself is negligible at 4 micron. Instead, Io's atmosphere appears in transmission either in the reflected component (two-way absorption) or in the volcanic component (one-way absorption). When modeling the (total) reflected components, we used a two-way air-mass µ=3.0. For a globally homogeneous atmosphere on a disk and no limb-darkening, the disk-averaged two-way transmittance would be expressed as $2E_3$ $(2\chi)$, where $\tau$ is the zenithal optical depth and $E_3$ is the third exponential integral (see e.g. Young et al. 2001). However, the geometry here is somewhat different as we are modeling spectra integrated along a 1-D slit. Based on numerical simulations for a slit much narrower than Io, we found that the equivalent air-mass is 2.87 for $\tau = 0.1$ (which is the typical optical depth in the core of the strongest lines), 3.07 for $\tau = 0.01$ and 3.13 in the limit of small optical depth. This justifies the choice of a mean $\tau=3$, and should not cause errors on the retrieved columns by more than ~5 %. When modeling the volcanic components, we used the precise air-mass corresponding to the location of the volcanic source on Io's disk during the observation.

Following the method of Lellouch et al. (2014, and references therein) for high-resolution spectroscopy, we directly modeled the Io spectra uncorrected for solar and telluric lines. "Monochromatic" spectra of Io's atmosphere (i.e. calculated with a $2\times10^{-4}$ cm$^{-1}$ step) were calculated along with model spectra of the Sun (Fiorenza and Formisano 2005) and of the Earth's atmosphere transmission (calculated for the appropriate $H_2O$ content and Io airmass during the observation, using LBLRTM[3] (Clough et al. 2005). When modeling the solar reflected part of Io's spectrum, these three components were multiplied, accounting for Io's topocentric and heliocentric velocities, before being convolved by the gaussian instrumental function. The width of the instrumental function was determined by fitting the telluric lines, and found to match expectations for a source filling the 0.5" or 0.8" slit (i.e. a spectral resolution R~40,000 or ~25,000). For the volcanic component, the same procedure was applied, except that the solar spectrum was not included in the product. Also in this case, a higher spectral resolution was determined (R~70,000), as expected for a more "point-like" (but still seeing-convolved) effective source size.

The de-convolution method of the solar reflected and volcanic components outlined above relies purely on the overall emission spatial profile. However the part of the emission attributed to the reflected component may still contain spatially unresolved thermal radiation from other hot spots, too weak to show a distinct profile. We have evidence for this situation from the fact that the modeled solar lines in the reflected component are often slightly too deep. We used this fact to determine a "thermal weight" (TW), to be applied to the model. In summary, the observed reflected component REF is modeled as:

---

[3] Available at rtweb.aer.com/lblrtm.html

$$REF = T_{earth} \; [(1-TW) \; T_{sol} \; T_{Io} \; (\chi=3) \; + \; TW \; T_{Io}(\chi=1.5)]$$

where $T_{earth}$, $T_{sol}$ and $T_{Io}$ are the transmission spectra due respectively to the Earth, the Sun, and Io's atmosphere. $T_{Io}$ is thus calculated for an air-mass $\chi=3$ for the reflected radiation and $\chi=1.5$ for the "unresolved" thermal radiation. TW is calculated near 3990 nm, where the most significant solar lines are present, and evaluated at other wavelengths by taking into account the general shape of the two components, i.e. the spectral behavior of Io's surface reflectivity (Fig. 2). In writing the above expression, Io's heliocentric and topocentric Doppler shifts have been omitted for the sake of simplicity, but are of course included in the model. Finally REF is convolved with the instrumental function with appropriate spectral resolution, and the data are compared to the model in terms of the line-to-continuum ratios.

## 6. Results
### 6.1 Reflected components

In a first step we modeled the nine reflected component spectra, focusing at this step on the "total" (i.e. integrated along the slit) spectrum. The 3980-4000 nm (2500-2513 cm$^{-1}$) range is the richest one in terms of $SO_2$ features that can be used to determine the $SO_2$ abundance. The first panel of Fig. 3 shows, on the example of spectrum 1, the comparison of the observed spectrum over 3985-3998 nm to models with different $SO_2$ column densities (0, 0.8, 1.3 and $2\times10^{17}$ cm$^{-2}$) and a single gas temperature (200 K). The sensitivity of the models to the $SO_2$ column indicates that we are able to determine the $SO_2$ abundance to within about ±15 % (in relation with the S/N of the data), if the temperature is known. However, the second panel of Fig. 3 shows the dependence of the spectrum to the gas temperature (120 K and 300 K), for a given $SO_2$ column. For that part of the spectrum, which mostly contains low-J lines, higher (resp. lower) gas temperatures result in lower (resp. higher) line depths, mimicking the effect of lower (higher) $SO_2$ columns. Discriminating between the different models can be achieved, by exploiting simultaneously regions of higher-J lines. There, line depths tend to increase with increasing temperature, as illustrated in the third panel of Fig. 3 (3966-3979 nm, i.e. 2513-2521 cm$^{-1}$) so that modeling the entire spectrum permits to some extent to disentangle temperature and abundance. In all three panels of Fig. 3, difference (observation minus model) spectra are shown. These residuals permit to distinguish spectral regions most sensitive to $SO_2$ and to determine best fits and associated uncertainties. Overall, the best fit for spectrum 1 is achieved for T = 200 K, with an estimated ± 40 K uncertainty. This temperature error bar adds another ±15 % uncertainty on the $SO_2$ abundance. Quadratically adding the two sources of uncertainty leads to a ±25 % final error bar on the $SO_2$ column (1.3±0.4 x 10$^{17}$ cm$^{-2}$).

The same procedure was applied to all nine spectra. Fig. 4 shows best fits of spectra 2-5 and B-E in the spectral regions most sensitive to $SO_2$ (3986-3998 nm, except for spectra 4 and 5 where it is 3978-3990 nm). Despite the weakness of the $SO_2$ absorptions, the very high S/N permits the unambiguous detection of $SO_2$ gas, even in the case of spectrum D which indicates a $3\times10^{16}$ cm$^{-2}$ column density only. Table 3 summarizes the determined $SO_2$ column and temperatures, as well as the value of the "thermal weight" that was included in the model to match the depth of the solar lines.

The measured $SO_2$ columns as a function of central meridian longitude (CML) are plotted in Fig. 5 (panel a) and compared with results from 19 µm spectroscopy (Spencer et al. 2005). Panels b) and d) of Fig. 5 show that to within error bars we cannot establish variations of the gas temperature with CML nor correlations with the gas density. Overall the determined temperatures are in the range 150-220 K. The weighted mean 167 K with a standard deviation of 14 K on the mean, and a 28 K standard deviation of data. Overall, we adopt a mean temperature of 170±20 K. As this is a surprising result in view of the colder temperatures obtained from 19-μm spectroscopy (<150 K, and especially 108±18 K on the anti-Jupiter hemisphere, see Tsang et al. 2012), we further demonstrate this visually in Fig. 6 on averaged spectra. To do so, we focus on the 6 reflected component spectra that have the same wavelength set-up and slit orientation (i.e. spectra 1, 3, B, C, D, E), and average them in two groups which have similar topocentric velocities, namely spectra (1, 3, E), and (B, C, D). Upon averaging, the spectra are re-aligned in the Io velocity frame, which permits to increase S/N on the $SO_2$ lines. Although this leads to a smearing of the solar and telluric lines (still somewhat reduced by the grouping of spectra with similar topocentric velocity), this is not per se a problem because one can apply exactly the same procedure to the synthetic spectra. The same method was successfully applied by Lellouch et al. (2015) when averaging high-resolution spectra of Pluto taken on different dates. Fig. 6a and Fig. 6b show the (1, 3, E) and (B, C, D) averages, respectively, compared to models with 220, 170, and 120 K. The first sub-panels of Fig. 6a and Fig. 6b shows the 3985-3998 nm, where, as an alternative presentation to Fig. 3, the $SO_2$ column is adjusted for an optimal match in this range. These models are then compared to data at 3966-3978 nm (middle panel) and 4011-4024 nm (bottom panel). For illustration of the sensitivity to temperature, observation-model residuals (multiplied by 2) are shown in Fig. 6a. The best overall fit is obtained for T = 170 K, where spectral contrasts as low as ~0.1 % are successfully reproduced. The associated best fit column density is 8.5 x $10^{16}$ cm$^{-2}$. Low temperatures such as T = 120 K can be excluded, as not producing enough absorption contrast in weak $SO_2$ features, especially over 3966-3978 nm. Results on the gas temperature and on the variation of the $SO_2$ column density with CML longitude will be further discussed below in Sections 6.3, 7.1 and 7.2.

Table 3. Derived $SO_2$ gas column density and temperature for the reflected components

| # | Central meridian longitude | Slit orientation | Thermal weight | $SO_2$ column (cm$^{-2}$) | Temperature (K) |
|---|---|---|---|---|---|
| 1 | 197-210 | Parallel | 0.15 | (1.3±0.4) x $10^{17}$ | 200 ± 40 |
| 2 | 212-225 | Perp. | 0.20 | (1.3±0.5) x $10^{17}$ | 200 ± 70 |
| 3 | 233-241 | Parallel | 0.075 | (1.0±0.3) x $10^{17}$ | 150 ± 40 |
| 4 | 123-136 | Parallel | 0.20 | (1.0±0.4) x $10^{17}$ | 180 ± 60 |
| 5 | 151-166 | Perp. | 0.20 | (0.5±0.25) x $10^{17}$ | 180 ± 80 |
| B | 61-79 | Parallel | 0.24 | (0.75±0.2) x $10^{17}$ | 180 ± 40 |
| C | 148-165 | Parallel | 0.25 | (0.9±0.25) x $10^{17}$ | 150 ± 30 |
| D | 284-302 | Parallel | 0.20 | (0.3±0.15) x $10^{17}$ | 220 ± 100 |
| E | 320-337 | Parallel | 0.20 | (0.75±0.2) x $10^{17}$ | 150± 30 |

6.2 Volcanic components

We next turned to modeling of the volcanic components extracted in spectra 1, 2, 3, D and E. Searching for $SO_2$ gas in the thermal component is considerably more difficult than in the reflected one, because (i) for equal column density, the one-way transmission leads to a loss of a factor of 2 in the line depths, and (ii) the continuum signal is lower, by a factor ~2, in the case of Pillan, and as much as a factor ~6 in the case of Loki (those numbers refer to the most diagnostic ~3990 nm spectral region). Fig. 7 shows the three volcanic component spectra at Pillan extracted on July 24, 2008 (obs. #1-3), along with best model fits. The spectra are displayed only in the most diagnostic 3985-3998 nm region, where clear $SO_2$ signatures are present. In fact, the $SO_2$ features are not detected unambiguously in the other spectral ranges, preventing a temperature determination to be performed. For this reason, a constant T = 170 K temperature was assumed when performing the fits. An alternate option would be to adopt the gas temperatures inferred from the associated solar reflected spectra, but this does not appear to us to be more justified. The best fit $SO_2$ columns are (6.5±0.15) x $10^{16}$, (8±2) x $10^{16}$, and (1.4±0.2) x $10^{17}$ cm$^{-2}$ for observations 1, 2 and 3 respectively, where error bars are associated to S/N limitations only (1-$\chi$ error bars and indicated). In these three spectra, Pillan was located 40.5°±6.5°, 25.5°±6.5°, and 7±4° sky East from the central meridian. For spectrum 3, the derived column is marginally higher but consistent with error bars with the slit-average value inferred from the corresponding solar reflected spectrum ((1.0±0.3) x $10^{17}$ cm$^{-2}$, see Table 2). This indicates that Pillan emits at most small additional $SO_2$ amounts (< 4x$10^{16}$ cm$^{-2}$, i.e. 30 %) above the "background" atmosphere. A bit surprisingly, the other two spectra indicate significantly lower $SO_2$ columns at Pillan, i.e. *less* than the mean atmosphere as measured in the associated reflected spectra. Although this result seems robust in terms of signal-to-noise (see Fig. 7), it is rather hard to interpret as a *depletion* of the local $SO_2$ atmosphere associated to the outburst. Assuming that the atmosphere above Pillan is simply unrelated to the presence of the hot spot (i.e., is of sublimation origin), the lower $SO_2$ columns measured in spectra 1 and 2 compared to spectrum 3 might be related to lower surface temperatures associated to the different local times (~9 am, ~10 am, and ~11:20 am respectively), although the effect seems large for a time excursion of less than 2.5 hours.

As mentioned above, a T = 170 K gas temperature is assumed in the above fits, while it is unconstrained in the volcanic spectra. Technically the Pillan spectra could also be fit with higher gas temperatures and higher $SO_2$ column densities. For example, the volcanic component of spectrum #1 (Pillan at 40° from Central Meridian) can also be fit with T = 500 K, $SO_2$ = 1.3x$10^{17}$ cm$^{-2}$. In this case, the column density is no longer lower than the "background" column as measured in the solar reflected component. However we do not regard this solution as physically plausible, because volcanic atmospheres are expected to be colder than sublimation atmospheres, due to the cooling associated to plume expansion and the short time available in plumes for solar heating.

Similarly, Fig. 8 shows the two volcanic spectra from Loki extracted from observations D and E, compared with models with and without $SO_2$ gas. Although the models reproduce the telluric absorptions rather well (especially for spectrum D), signal-to-noise limitations associated with the much lower thermal flux than at Pillan (see Fig. 2) do not permit us to unambiguously identify $SO_2$ and an upper limit of 1.5x$10^{17}$ cm$^{-2}$ is estimated. Since the $SO_2$ column in the associated solar reflected component is of order ~0.5x$10^{17}$ cm$^{-2}$, we conclude that the maximum $SO_2$ amount produced by Loki above the local background is ~1x$10^{17}$ cm$^{-2}$. This still leaves the possibility that Loki is a major contributor to the atmosphere near the L~300 longitude.

## 6.3 Spatially resolved spectra: latitudinal and diurnal variations

Finally, we further exploited the mapping capability of CRIRES to search for variations of the spectra along the slit. As indicated before, the typical seeing-limited spatial resolution of the data is about 0.2", which would in principle provide ~6 independent points along the slit across Io's diameter. However, this cannot be achieved in practice because of S/N limitations.

<u>Latitudinal distribution:</u> Data for which the slit was aligned parallel to the polar axis (#2 and #5) allow us to study the distribution of Io's atmosphere with latitude. We considered mostly dataset #2, for which the $SO_2$ columns are larger. Spectra from this dataset were combined to produce two average spectra, corresponding respectively to (i) a "low-latitude" (-40° to +40°) spectrum (corresponding to the central 9 pixels), and (ii) a "high-latitude" (polewards of ±40°) spectrum, associated with the remaining pixels. These two spectral averages were modeled by considering different $SO_2$ distributions, namely (a) uniform distribution (b) the "modified latitude dependence" case of Spencer et al. (2005, see their Fig. 7a), tuned to match HST spectroscopic results from Jessup et al. 2004 (wherein the effects of time of day and latitude were mixed) (c) an intermediate distribution, characterized by Gaussian fall-off of the density away from equator, with the density halved at ±40° (i.e. a Gaussian with HWHM = 40°). These three distributions (relevant to CML = 212-225) are shown in Fig. 9a. For each of them, synthetic spectra were calculated on a fine latitude grid, accounting for the exact local air-mass. The spectra were then convolved spatially by a 1-dimension Gaussian function with FWHM = 0.2", mimicking the seeing effects, and the spectra convolved in this manner were then averaged in the same two groups as the observations. In each case, a 200 K gas temperature (Table 2) was adopted and the equatorial density was adjusted to provide the optimum match to the low-latitude spectrum. Results and comparison to the data are shown in Fig. 10. As can be expected, the "low latitude" spectrum (top panel) provides no discrimination between the three models. In contrast, the "high latitude" spectrum (bottom panel) clearly rules out the uniform with latitude distribution case, which would lead to much too deep $SO_2$ features (as can also be seen from the observation-model residuals). Conversely, the latitude distribution adopted by Spencer et al. (2005) tends to underestimate the $SO_2$ lines and the intermediate distribution (red curve of Fig. 9a), associated to an equatorial column density of 2 x $10^{17}$ cm$^{-2}$, provides the overall best match. Fig. 9b shows diurnal distributions of $SO_2$, to be discussed later.

Similar work was performed for dataset #5, averaging this time data at latitudes equator-ward and poleward of ±30°). Again, the uniform with latitude case could be excluded, but in this case, the data quality was insufficient to discriminate between the other two models. These two models have an equatorial column density of 0.75 x $10^{17}$ cm$^{-2}$.

We conclude that our data confirm the evidence for a latitudinally-restricted atmosphere found in previous datasets, and our best assessment is that of a factor-of-2 decrease in the $SO_2$ columns from 0° to ±40°. We note also that the equatorial column densities inferred with this distribution are ~1.5 times larger than from those from the corresponding slit-averaged spectra, assuming constant $SO_2$ along the slit (see Table 2). Therefore, for comparison with the columns derived from all other spectra (with the slit parallel to equator), it is appropriate to correct the $SO_2$ columns for observations #2 and #5 in Table 2 upwards by a factor 1.5. These corrected columns are also shown in Fig. 5 (panels c and d).

<u>Diurnal variations:</u> We used a similar approach, but a further complication is that diurnal variations are intermixed with longitudinal variability in producing the variability of the $SO_2$ spectrum along the slit. To minimize the effect of geographical variations, we first build a "grand-average" of all 6 observations that have the same wavelength set-up and slit orientation (i.e. data 1, 3, B, C, D, E). However, unlike was done previously (Section 6a) when determining the gas temperature on averaged spectra, we now retain the spatially-resolved information. In practice, for each of the above six observations, we extract three reflected component spectra: (i) a "noon spectrum" that corresponds to the 5 central pixels and samples longitudes within ±20° from central meridian (i.e. local times ±1h20 min from noon), and (ii) a "morning spectrum", integrated over all pixels that are sky-east from the noon spectrum (local times from ~6 am to 10:40 am) (iii) an "evening spectrum", similarly corresponding to local times from ~1:20 pm to 6 pm. [Note that these local times are approximate because the phase angles are not strictly equal to zero. However, except for spectrum D for which the sub-observer point is shifted by 40 min after local noon, these effects are of order 5-15 minutes only]. Once extracted, these spectra are averaged over the 6 observations after being re-aligned in the Io velocity frame, ending up in one "noon", one "morning" and one "evening" spectra.

The spectra are then modeled for different assumptions on the diurnal distribution of $SO_2$ gas, assuming a fixed temperature of 170 K. The approach is the same as above for the latitudinal distribution (in particular to handle seeing effects), the only difference being that any given model has to be calculated for the 6 individual datasets, re-aligned, and co-added for comparison with the data.

Fig. 11 shows the three extracted spectra and model fits for five assumed diurnal distributions of $SO_2$ (shown in Fig. 9b). One of them (green line) is uniform with local time, and the others specify a simple Gaussian fall-off of the $SO_2$ column density away from disk-center – confused with local noon – with the density halved at some value (HWHM), to which a finite "background" value may be added. Starting with pure Gaussian distributions, the red and blue models have HWHMs of ±40° (i.e. ±2.7 hours from noon) and ±80° (i.e. ±5.3 hours from noon). The column density at noon in each case is adjusted so as to match the "noon" spectrum. For a uniform distribution of $SO_2$ with local time (green lines in Fig. 11), this requires a $SO_2$ column density of $1.1 \times 10^{17}$ cm$^{-2}$ at noon. This is 30 % higher than the value required to fit the "total" (mean) reflected spectrum ($8.5 \times 10^{16}$ cm$^{-2}$, see Fig. 6), which is already an indication that the $SO_2$ columns decrease away from noon. As a matter of fact, the second and third panels of Fig. 11 show that this uniform case over-predicts the $SO_2$ absorptions in the "morning" and "evening" spectra. The result is especially clear for the morning spectrum and somewhat more marginal for the evening data. Better fits are achieved for the other two described cases, with the HWHM = 40° (resp. 80°) case providing an optimum match of the morning (resp. evening) spectrum. We conclude that the data show evidence for a moderate diurnal variability of $SO_2$, with a suggestion that the atmosphere is more extended on the evening vs. morning side, probably resulting from a surface temperature distribution affected by thermal inertia effects. As the latter push the temperature maxima into early-to-mid afternoon hours, we recognize that adopting a $SO_2$ distribution with maximum at noon is not necessarily justified, but more complex models may not be warranted given S/N limitations.

Tsang et al. (2012, 2013b) find that the ensemble of the 19-σm data over 2001-2013 indicates the co-existence of sublimation and volcanic atmospheres, the latter contributing to (assumed independent with local time, longitude and epoch) a 0.55–0.75 x $10^{17}$ cm$^{-2}$ column. To test this scenario we consider two more diurnal distributions in the form of a Gaussian superimposed to a constant (0.55-0.6) x $10^{17}$ atmosphere. As the addition of this term reduces the diurnal contrast, fitting the "morning" and "evening" spectra requires a steeper diurnal variation of the Gaussian component. For the evening spectrum, an HWHM of ±60° is now required (purple curve). For the morning, an HWHM of at most ±20° is needed, and the fit tends to over-predict the line-depths, suggesting that the constant (volcanic) contribution may be somewhat too large.

Fig. 12 shows our final best estimate of the $SO_2$ column densities at noon as a function of CML. Based on the above results on the latitudinal and diurnal variations, these are obtained by increasing the measured slit-averaged columns by 30 % (resp. 50 %) for slit orientation parallel (resp. perpendicular to equator). Maximum noon column densities of ~2 x $10^{17}$ cm$^{-2}$ are indicated.

## 7. Discussion

### 7.1 $SO_2$ distribution and volcanic vs. sublimation atmosphere

Volcanic thermal emission – caused by an outburst at Pillan in 2008 and due to moderate activity at Loki 2010 – can be identified in the data from its peaked spatial signature and separated from the dominant, spatially quasi-uniform component that is predominantly due to the solar reflected component (with minor thermal contribution). The $SO_2$ atmosphere can thus in principle be investigated separately in the two components. Most of our results are obtained in the solar reflected component, providing new results on the mean atmospheric density, and its latitudinal, longitudinal, and diurnal variability.

The spatial distribution of Io's atmosphere has been addressed in many past works, but so far the more explicit results have been obtained from mid-UV spectroscopy, Ly α imaging, and thermal IR spectroscopy. Lyα images taken over 1997-2001 (Feldman et al. 2000, Strobel and Wolven 2001, Feaga et al. 2009) and interpreted in terms of $SO_2$ continuum absorption have provided the first 2-D maps of the $SO_2$ atmosphere, showing a denser and somewhat more extended atmosphere on Io's anti-jovian hemisphere, with maximum densities of ~5 x $10^{16}$ cm$^{-2}$ near L=140° longitude on the equator, minimum equatorial densities of ~1.5 x $10^{16}$ over L=240°-360°, and relatively sharp latitudinal boundaries (at about ±45° on the anti-jovian side and ±30° on the sub-jovian side). However, the longitudinal distribution and the absolute columns from Feaga et al. (2009) are only qualitatively consistent with those inferred from Spencer et al. (2005) from disk-averaged but rotationally-resolved 19-µm data over 2001-2004. The latter also point to more gas on the anti-jovian hemisphere, but the peak density is ~1.4 x $10^{17}$ cm$^{-2}$ (averaged over 2001-2004) and occurs near L = 220°, with a factor-of-10 contrast with densities at L = 290°-330°. Based on subsequent 19-µm data which show the seasonal evolution of Io's atmosphere, Tsang et al. (2012, 2013b) find that an equatorial maximum column density as low as ~5 x $10^{16}$ cm$^{-2}$ rather occurs near aphelion (e.g. March 2005).

Fig. 5 (panel c) shows the equatorial densities inferred in this work; the corresponding sub-solar densities as a function of CML are shown in Fig. 12. For the epoch from mid-2008 to mid-2010, we determine a maximum equatorial density of ~1.5x$10^{17}$ cm$^{-2}$ at central meridian longitude L = 200-220 and a minimum of ~3x$10^{16}$ cm$^{-2}$ at L = 285°-300°. Given that in terms of heliocentric distance, the 2008-2010 period is similar to 2001-2003, the maximum value we measure and its phase are much more consistent with the Spencer et al. (2005, their Fig. 17) results than with Feaga et al. (2009). Our peak values are also in general agreement with HST/STIS spectroscopic results from observations in late 2001 (Jessup et al. 2004, their Fig. 8) and 2010-2012 (Jessup and Spencer 2015; their Fig. 13), and with HST/COS data from 2010 (Tsang et al. 2013a); in particular Jessup and Spencer (2015) determine maximum zero-latitude $SO_2$ column densities of ~ (2.0-2.5) x $10^{17}$ cm$^{-2}$, occurring at L =170. The longitudinal contrast we measure is however not as strong as from the 19-µm data. Differences are especially noticeable in the sub-jovian hemisphere, where our densities are larger and show a significant increase of the $SO_2$ columns from L = 290° to L=330°. Within the accuracy of the measurements, we cannot distinguish any evolution of the atmospheric density from mid-2008 to mid-2010.

Our measurements represent the first spatially-resolved infrared observation of the $SO_2$ atmosphere (except for Voyager 1), providing information on both its latitudinal and diurnal behavior. The decrease of the $SO_2$ columns with latitude is apparent. The typical latitudinal extent of the atmosphere is found to be ±40° at half-maximum, and the equatorial column density is ~1.5 times stronger than the geometrically-averaged value over all latitudes. Explicitly testing the latitudinal distribution adopted by Spencer et al. (2005) (based on matching the HST spectroscopic results from Jessup et al. 2004) indicates that it decreases too sharply with latitude to fit our data, at least at L ~220°. Similarly, a factor-of-10 decrease of the $SO_2$ columns between the equator and about ±40° latitude, as inferred from the Lyα data (Feaga et al. 2009, their Fig. 15) would not produce enough absorption in our "high-latitude" (i.e. polewards of ±40°) spectra. This inconsistency, as well as that on the absolute columns, suggests that the interpretation of the Lyα data is affected by issues with the assumed surface reflectance at this wavelength and its latitude dependence. Moullet et al. (2008) also find that the absolute column densities reported in Feaga et al. (2009) are too low to match their mm-wave measurements.

Studying the diurnal variability is complicated by the fact that spectral variations along the equator are also affected by geographical variations, and that the weakness of the $SO_2$ absorptions imposes significant averaging along the slit. Nonetheless, by combining datasets at six different longitudes, we were able to extract a diurnal signature in the equatorial atmosphere. For models with simple exponential decay of the density away from local noon, factor of 2 lower densities occur ~40° (2.7 hours) before and ~80° (5.3 hours) after noon. When allowance is made for a volcanic component of the magnitude inferred by Tsang et al. (2013b), the required HWHM are reduced to <20° (1.3 hours) in the morning ~60° (4 hours) in the evening. We checked that this apparent diurnal variability is unlikely to be caused by the geographical variations. For this, assuming the longitudinal distribution of $SO_2$ of Spencer et al. (2005, reproduced here in Fig. 5), we calculated the expected $SO_2$ column averaged over the six CML longitudes of observation. This average was calculated at noon (i.e. using a longitude equal to the CML in the observation), and then for "morning" and "evening" conditions, for which a longitude displaced by +60° or -60° from the observed CML was used. With that, the expected "morning" average is ~30 % less than the "noon" average, and the "evening" one is 10 % larger. Note that these numbers are probably upper limits, as the Spencer et al. (2005) distribution has sharper contrasts than measured here (Fig. 5) and from HST/UV imaging and spectroscopy (see Fig. 14b of Jessup and Spencer 2012). This confirms that the method minimizes the effect of the uneven geographical distribution. In particular, we measure lower $SO_2$ amounts in the evening average compared to noon, while ~10 % larger would be expected if purely geographical effects were at work.

The diurnal variation is evidence of a sublimation-supported component sensitive to surface temperature, where a factor-of-2 pressure variation about ~2 nbar (1.1 x $10^{17}$ cm$^{-2}$) is driven by a ±2 K variation about a sublimation temperature of 116.2 K (using the Wagman (1979) equation). From Galileo/PPR data, Rathbun et al. (2004) find a peak dayside temperature of ~130 K and a best-fit thermal inertia TI = 70 W m$^{-2}$ s$^{-1/2}$ K$^{-1}$ (MKSA) units (or a two-component model with TI = 30 and TI = 100 MKSA). If SO$_2$ pressure control occurs at the surface, this temperature is much too high to account for the observed columns. From the evolution of the SO$_2$ columns with heliocentric distance, Tsang et al. (2012, 2013b) find a frost thermal inertia in the 100-800 MKSA range, and as pointed out by the authors, even these high values allow for significant diurnal temperature variability. For example, for an heliocentric distance $R_h$ = 5.07 AU, representative of our measurements, a Bond albedo of 0.55 and TI = 400, the peak dayside temperature is 116.2 K, and the temperatures at -40° and +80° longitude from noon are 111.0 and 113.6 K respectively, i.e. 4 K and 1.4 K colder than the noon temperature (115.0 K). Although not perfect, this is the order of magnitude suggested by the diurnal variations seen in our data. The agreement improves (degrades) for the morning (evening) spectrum when allowance is made for the volcanic component, making it difficult for us to further constrain the thermal inertia. At any rate, this issue is also hampered by the possibility of sub-surface control of the atmosphere (Matson and Nash 1983, Kerton et al. 1996), as there is an almost complete degeneracy between thermal inertia and depth into the surface in dayside temperature curves. Finally, we note that the apparent lack of diurnal variation in atmospheric density inferred from the Lyman-α images can in fact be consistent with sublimation support by high thermal inertia frost (Spencer et al. 2005).

From HST/STIS measurements over Dec. 2010- Jan. 2012, Jessup and Spencer (2015) searched for diurnal variations of the atmosphere at a given longitude and found it to be at best marginal over a 50° (3.3 hours) interval (if expressed in terms of a different sublimation temperature, the difference in the SO$_2$ equatorial column at ~10 am and at ~1 pm would correspond to a mere -0.6 K temperature decrease from mid-morning to early afternoon). In any case, diurnal variations were found to be dwarfed by longitudinal variations at a given local time. For example, Jessup and Spencer (2015) found a ~$10^{17}$ cm$^{-2}$ difference between the columns observed at L = 200 and L = 250, for equal local time. This is equivalent to a ~2.4 K difference in the sublimation temperature at noon between L = 200 and L = 250. At face value the lack of diurnal variability seems to be inconsistent with our above finding and further investigations are needed. Note that the HST/STIS measurements sample a variety of latitudes over ~30°S-30°N, so that an intermediary required step is to correct the measurements to a common "zero latitude" (which is done by using a cos$^{1/4}$ (latitude) dependence). The magnitude of this correction is not small (equivalent to ~2 K at latitude = 20°), which introduces additional complexity.

Our data provide more limited results in the volcanic component, due to the factor-of-2 (Pillan) or factors of 6-8 (Loki) lower fluxes. The Loki spectra do not have enough S/N for us to detect the SO$_2$ signature, but can still be used to set an upper limit of 1.5x$10^{17}$ cm$^{-2}$ for the SO$_2$ column density above Loki, indicating a maximum contribution from a possible volcanic atmosphere above the local background of ~1x$10^{17}$ cm$^{-2}$. This is much smaller than the original figure reported from the Voyager 1/IRIS detection (about 5x$10^{18}$ cm$^{-2}$, Pearl et al. 1979) and even smaller than the (2.5-21) x $10^{17}$ cm$^{-2}$ range from the reanalysis of Lellouch et al. (1992). This difference can probably be attributed to the fact that the IRIS FOV included a plume over which the SO$_2$ concentration may have been enhanced, and which has not been observed to be active since then.

There are no direct measurements of SO$_2$ densities in the Pillan plume itself, although Jessup et al. (2007) have speculated that lava flows NE of Pillan were the source of an active plume, detected at ~6 S, 264 W latitude (i.e. 170 km away) in February 2003 and containing ~4 x 10$^{16}$ cm$^{-2}$ of SO$_2$ (and ~3 times less S$_2$). Jessup and Spencer (2012) further interpreted the 0.26 μm brightness of the 1997 Pillan plume as being attenuated by a SO$_2$ gas component with column density (3-6) x 10$^{18}$ cm$^{-2}$. Our volcanic spectrum at Pillan does show the signature of SO$_2$ gas, but the amount is not significantly higher than in the reflected spectra taken simultaneously, indicating that Pillan emits at most weak (< 4 x 10$^{16}$ cm$^{-2}$) extra SO2 above the "background" atmosphere. This figure is similar to values reported previously for other plumes, namely Pele (7 x 10$^{16}$ cm$^{-2}$ in tangential viewing, or ~2 times less when converted to a vertical column, Spencer et al. 2000) on Io's limb (i.e. terminator), and Prometheus, where a local enhancement of 5 x 10$^{16}$ cm$^{-2}$ above mean equatorial values was measured (Jessup et al. 2004). The latter comparison is especially relevant as the Pillan plume is classified as Prometheus-type (Geissler and McMillan 2008). Finally, detecting the Pillan emission in three spectra over a 2.5 hr interval (local time ~9 am, ~10 am and ~11.3 am), we also tentatively find variations of the SO$_2$ atmosphere above the hot spot with local time, in qualitative agreement with our inference for diurnal variability as discussed above.

### 7.2 Gas temperature

Determining the mean temperature of Io's atmosphere (not to mention the vertical temperature structure) has proven to be one of the most difficult tasks. Millimeter-wave observations are highly sensitive to gas temperature (which impacts both line contrasts and line widths), but as outlined in the introduction, the large body of available data has given conflicting results on the mean gas temperature. Initial results from 1990-1994, in which the line widths were understood as due to a combination of thermal broadening and saturation effects, pointed to very high (500 K-600 K on trailing side, 250-400 K on leading side) values (see Lellouch et al. 1992, Lellouch 1996), somewhat relaxed (200-400 K) in observations from 1999-2002 (see McGrath et al. 2004, Lellouch et al. 2007). These high values were put into question when two alternative interpretations of the line widths were proposed: (i) that they represent velocity dispersion within volcanic plumes (Ballester et al. 1994, Lellouch 1996), and (ii) that they represent planetary-scale velocities, direct evidence for which was obtained from interferometric measurements resolving the Io disk (Moullet et al. 2008). With the latest interpretation, gas temperatures determined from IRAM-30m observations in 1999 and IRAM-PdBI in 2005 span the range 130−210 K (Moullet et al. 2008), although more recent observations (from APEX) suggest again high temperatures (Moullet et al. 2013). One point to stress here is that most of the observed mm-lines have low energy levels (< 110 cm$^{-1}$), so that the reported temperatures are not equivalent of a rotational diagram analysis. An exception is the January 2002 observations from Lellouch et al. (2003), which included lines with energy levels up to 404 cm$^{-1}$ (580 K), and which indicated a temperature of 180±60 K.

In general, UV observations are more weakly diagnostic of temperature, which can still be constrained from subtle skewness effects in the spectrum. Inferred (or assumed) temperatures (Ballester et al. 1994, McGrath et al. 2000, Spencer et al. 2000, Jessup et al. 2004, 2007, Tsang et al. 2013a, Jessup and Spencer 2015) vary from 110 K to 500 K. The most recent and highest quality observations tend to favor the lower temperature range (150-250 K from Jessup et al. 2004, 100-200 K from Tsang et al. 2013a and Jessup and Spencer 2015).

The detection of the μ$_2$ band of SO$_2$ at 19 νm has provided new temperature constraints. Using a non-LTE model, and based on the maximum observed absorption depth for the strongest 530.42 cm$^{-1}$ feature, Spencer et al. (2005) found that the gas kinetic temperature has to be less than ~140-150 K, at least on the anti-Jupiter hemisphere. This constraint may still be considered as model-dependent since it depends on the accuracy of the non-LTE model (i.e. its ability to properly calculate vibrational temperatures as a function of gas kinetic and surface temperature). Tsang et al. (2012) expanded on this result by fitting the entire 529.0-530.7 cm$^{-1}$ range, and found that low temperatures (< 150 K) are further required to match the overall spectral shape, in particular a 530.43 cm$^{-1}$ shoulder in the wing of the main SO$_2$ line (their Fig. 8). For the L = 90-220 range, they inferred an extremely cold mean value of 108 ±18 K over 2001-2010.

The best-fit temperature values for all of our spectra are in the range 150- 220 K. The mean value is T = 170±20 K and we have shown that T = 120 K (and a fortiori lower temperatures) is inconsistent with the mean 4.0 µm spectrum. The contradiction with Tsang et al. (2012) is puzzling as in a sense both measurements are direct rotational temperature measurements (particularly true in our case, since all lines are optically thin). A difference between the two experiments is the fact that one samples the reflected component and the other the thermal component, but there is no obvious reason why this should lead to two apparently different temperatures. In our observations, the evidence for diurnal variations and the lack of additional gas above Pillan point to the fact that at least a fraction of the sampled atmosphere is sublimation-supported. A possible explanation could be that for some reason the thermal data would be more weighted towards volcanic gas, which is cold due to plume expansion except in the region of the canopy shock (Zhang et al. 2003). However, this is not obviously consistent with the inference, based on the atmospheric evolution with heliocentric distance as seen in the 19 µm data, that the contribution of volcanic and sublimation components are typically comparable (Tsang et al. 2012, 2013b). For example, at longitude L=180, the preferred scenario of Tsang et al. 2013b has a volcanic component of ~6.5 x 10$^{16}$ cm$^{-2}$ and a sub-solar sublimation contribution of ~2 x 10$^{16}$ cm$^{-2}$ at aphelion (e.g. March 2005) and ~1.3 x 10$^{17}$ cm$^{-2}$ at perihelion (e.g. March 2011). We note finally that our temperature is consistent with the determination based on the Jan. 2002 millimeter measurements of Lellouch et al. (2003), i.e. also in the thermal range, that included high- and low- lower energy lines, so that a correlation between the measured gas temperature and the measurement wavelength does not seem to be supported. A final, speculative, hypothesis would be that rotational temperatures are not directly representative of gas kinetic temperatures, and that differential non rotational-LTE effects occur between the µ$_2$ and ν$_1$ + ν$_3$ bands, but this does not seem plausible for surface pressures of ~1 nbar (Lellouch et al. 1992, Spencer et al. 2005). As a concluding point, we note that our mean value of T = 170±20 K is plausibly consistent with expectations based on radiative-conductive model of a hydrostatic atmosphere in equilibrium with solar input (Strobel et al. 1994). These authors considered two atmospheric end-members having respectively surface pressures of 130 nbar and 0.3 nbar, bracketing observations (~3 nbar). While the "high"-pressure case leads to strong IR cooling and attendant cold temperatures (<130 K) in the first ~40 km, the low-pressure scenario has temperatures that quickly ramp to ~170 K at 10 km (i.e. ~1 scale height)

8. **Summary**

From VLT/CRIRES 1-D slit spectroscopic observations acquired in 2008 and 2010, sampling a variety of observing central meridian longitudes, we have reported the first detection of $SO_2$ in Io's atmosphere in the $\nu_1+\nu_3$ band at 4.0 νm, and obtained the following characterization:

- Io's atmosphere shows unambiguous longitudinal asymmetry, as reported in previous works. For 2008-2010, we determine a maximum equatorial density of ~$1.5 \times 10^{17}$ cm$^{-2}$ at central meridian longitude L = 200-220 and a minimum of ~$3 \times 10^{16}$ cm$^{-2}$ at L = 285°-300°. Accounting for the heliocentric dependence of the anti-jovian atmosphere reported by Tsang et al. (2012, 2013b), the maximum value of the density and its phase are generally consistent with earlier inferences from thermal IR and HST spectroscopy, but much less so with HST/Ly α imaging. Nonetheless, the longitudinal pattern we determine somewhat differs from inferences based on HST and thermal IR measurements. Within the accuracy of the measurements, we cannot distinguish an evolution of the atmospheric density from mid-2008 to mid-2010.

- The decrease of the $SO_2$ columns with latitude is apparent, albeit less sharp than indicated from HST mid-UV spectroscopy. The typical latitudinal extent of the atmosphere is found to be ±40° at half-maximum.

- The data show evidence for a diurnal variation of the equatorial atmosphere, with factor of 2 lower densities ~40° (2.7 hours) before and ~80° (5.3 hours) after local noon. This diurnal variation is evidence of a sublimation-supported component, with frost temperature variations of order ~2 K on these time intervals.

- The best-fit gas temperatures range from 150 to 220 K, with a weighted mean value of 170±20 K. This value should represent the column-weighted mean temperature of Io's atmosphere. The contradiction with the colder (~110 K) temperatures measured from the $\mu_2$ band in the thermal infrared remains to be understood.

- Finally, in spite of the clear thermal emission signal due to Pillan (in outburst in July 2008) and Loki (at typical levels in 2010), no detectable enhancements in the $SO_2$ atmosphere above these volcanic regions are found, with an upper limit of $4 \times 10^{16}$ cm$^{-2}$ at Pillan and $1 \times 10^{17}$ cm$^{-2}$ at Loki for any additional localized gas above the background atmosphere. Since the background density is only ~$5 \times 10^{16}$ cm$^{-2}$ at the Loki longitude, we cannot exclude however that the atmosphere near L ~ 300 be strongly contributed by volcanic activity.

**Acknowledgements.** Based on ESO observation programs 081.C-0564 and 385.C-0372. We are indebted to ESO personal for service mode observations in 2010. We thank John Spencer and an anonymous referee for constructive reviews.

# Figure Captions

**Fig.1.** Evidence for and extraction of thermal emission component in spectral images 1, 2, 3, D, E. (Left): Spectral images**.** The x-axis is the spectral dimension of the first array (1024 pixels covering 3958-3978 nm). The y-axis shows a portion of the spatial dimension, restricted to 200 spatial pixels centered on Io. The y-axis is oriented from sky East to West except in image 2 where it is oriented from North to South. (Right). Associated spatial profiles of the measured radiance at one particular wavelength (solid line), and their fits (see text for details). Dashed line: thermal component; dashed-dotted line: solar reflected; dotted line: sum of the two components. The reference pixel (pixel 0) is taken at the maximum of the total emission.

**Fig.2**. Extracted solar reflected and thermal spectra for observations 1, E and D (from top to bottom) See text for details on the extraction and calibration method. The grey line at the top shows the solar spectrum (arbitrary units).

**Fig.3.** Models of the reflected component in observation 1. Black lines: observations, expressed as transmission relative to local continuum vs. wavelength (in the Earth frame). Light blue lines: models without any contribution from Io's $SO_2$ atmosphere. Other colored lines refer to models including $SO_2$ with various column densities and temperatures, as indicated. The first two panels show separately the effects of column density and temperature in the 3985-3998 nm range. The third panel shows the effect of temperature in the 3966-3979 nm range. These models illustrate how temperature and abundance effects can be disentangled. Observations are vertically shifted with respect to models for improving plot clarity. Residuals between observations and models are shown.

**Fig.4.** Best fit spectra for the eight other reflected component spectra (observations 2-5 and B-E). Black lines: observations. Red lines: best fit. Green lines: models without $SO_2$ gas. Note the lower spectral resolution in the case of spectrum #2 (25,000 instead of 40,000).

**Fig.5.** a): Mean $SO_2$ column density measured over the slit measured in the reflected component as a function of Io's central meridian longitude (CML). b): Measured gas temperature vs. CML. Red symbols: observations from 2008 with the slit parallel to equator (1, 3 and 4). Blue symbols: observations from 2008 with the slit perpendicular to equator (2, 5). Green symbols: observations from 2010. c): same as a) but the column density for observations 2 & 5 has been rescaled by a factor 1.5 for more direct comparison with the equatorial column densities measured in other observations (see Section 6.3). The thin lines in the column density plots represent the equatorial column densities inferred from the 2001-2004 19-µm data of Spencer et al. (2005). d): equatorial column densities as a function of measured gas temperature.

**Fig.6**. Fit of the reflected component in two spectral averages. **a)** (1, 3, E) average **b)** (B, C, D) average. These averages are constructed by re-aligning spectra in the Io velocity frame (see text). Black lines: observations. Light blue: models without $SO_2$ gas. Red, green and dark blue lines show models with gas temperatures of 220, 170 and 120 K, respectively. Top panel: 3985-3998 nm. For each temperature, the $SO_2$ column is adjusted for optimal match in this range. These models are then compared to data at 3966-3978 nm (middle panel) and 4011-4024 nm (bottom panel). The best overall fit is obtained for T = 170 K.

**Fig.7.** Fit of the three (Pillan) volcanic spectra from July 24, 2008 in the 3985-3998 nm range. (Top). Extracted spectrum from observation 3, with Pillan 7±4° sky East from the central meridian. (Middle). Same from observation 1, with Pillan 40.5±6.5° sky East from CM. (Bottom). Same from observation 2, with Pillan 25.5±6.5° sky East from CM. Note how different the spectra are from the reflected spectra (absence of solar lines and sharper telluric features due to the spatially smaller effective source in the thermal spectra). Spectra are compared with models with a fixed temperature (T = 170 K). The 1-µ error bars and indicated.

**Fig. 8.** Two volcanic spectra (Loki) from July 27, 2010 (spectrum D, top) and September 27, 2010 (spectrum E, bottom) in the 3985-3998 nm range, are compared with models with and without $SO_2$ gas. Although the models reproduce the telluric absorptions rather well (especially for spectrum D), $SO_2$ is not unambiguously identified and an upper limit of $1.5 \times 10^{17}$ cm$^{-2}$ is estimated. The 1-σ error bars and indicated.

**Fig. 9.** Latitude (a) and local time (b) distributions of $SO_2$ considered in this work. (a) Three latitude distributions for the analysis of dataset #2 (corresponding to CML = 212-225). Green: uniform. Blue: "modified latitude dependence" (see Spencer et al., 2005). Red: intermediate distribution, having a gaussian fall-off from equator with a factor-of-2 drop-off (HWHM) at ±40°. This latter distribution provides the best fit (Fig. 10). (b) Five distributions of the equatorial $SO_2$ with local time (expressed as the longitude difference from noon), relevant to the analysis of the "grand-average" dataset combining the 1, 3, B, C, D, and E observations (see Section 6.3). Green: uniform distribution. Red and blue: Gaussian distributions with HWHM= ±40° and ±80°. Grey and purple: sum of a constant (volcanic) component of (0.55 – 0.6) x $10^{16}$ cm$^{-2}$ and a Gaussian with HWHM= ±20° (grey) and HWHM= ±60° (purple).

**Fig. 10.** "Low-latitude" (-40° to +40°) (top) and "high-latitude" (poleward of ±40°) (bottom) spectra corresponding to dataset #2. They are compared with models with the three latitude distributions of Fig. 9a. The equatorial column densities in Fig. 9a are adjusted for optimum match of the low-latitude spectrum, where the three models yield virtually identical fits. Residuals between observations and models are shown.

**Fig. 11.** "Noon", "morning" and "evening" spectra constructed from datasets 1, 3, B, C, D and E (see Section 6.3) are compared to models with different diurnal distributions of $SO_2$ (shown in Fig. 9b) Light blue: No $SO_2$. Green: uniform distribution with local time. Dark blue: Gaussian distribution centered at disk center (~local noon) and with HWHM = 80°. Red lines: same as blue, but HWHM = 40°. Grey: sum of a Gaussian distribution with HWHM = 20°

and a volcanic component of $6 \times 10^{16}$ cm$^{-2}$. Purple: sum of a Gaussian distribution with HWHM = 60° and a volcanic component of $5.5 \times 10^{16}$ cm$^{-2}$. The SO$_2$ column at noon is always adjusted (at $1.1 \times 10^{17}$ cm$^{-2}$ or $1.2 \times 10^{17}$ cm$^{-2}$) in order to match the "Noon" spectrum (where all models therefore coincide). The gas temperature is taken as 170 K. Residuals between observations and models are shown.

**Fig. 12.** Sub-solar column densities (measured in reflected component) vs. Io's central meridian longitude. They result from the upwards correction of the measurements shown in panel a) of Fig. 5, by a factor 1.3 (resp. 1.5) for measurements with the slit parallel (resp. perpendicular) to equator. See Section 6.3 for details.

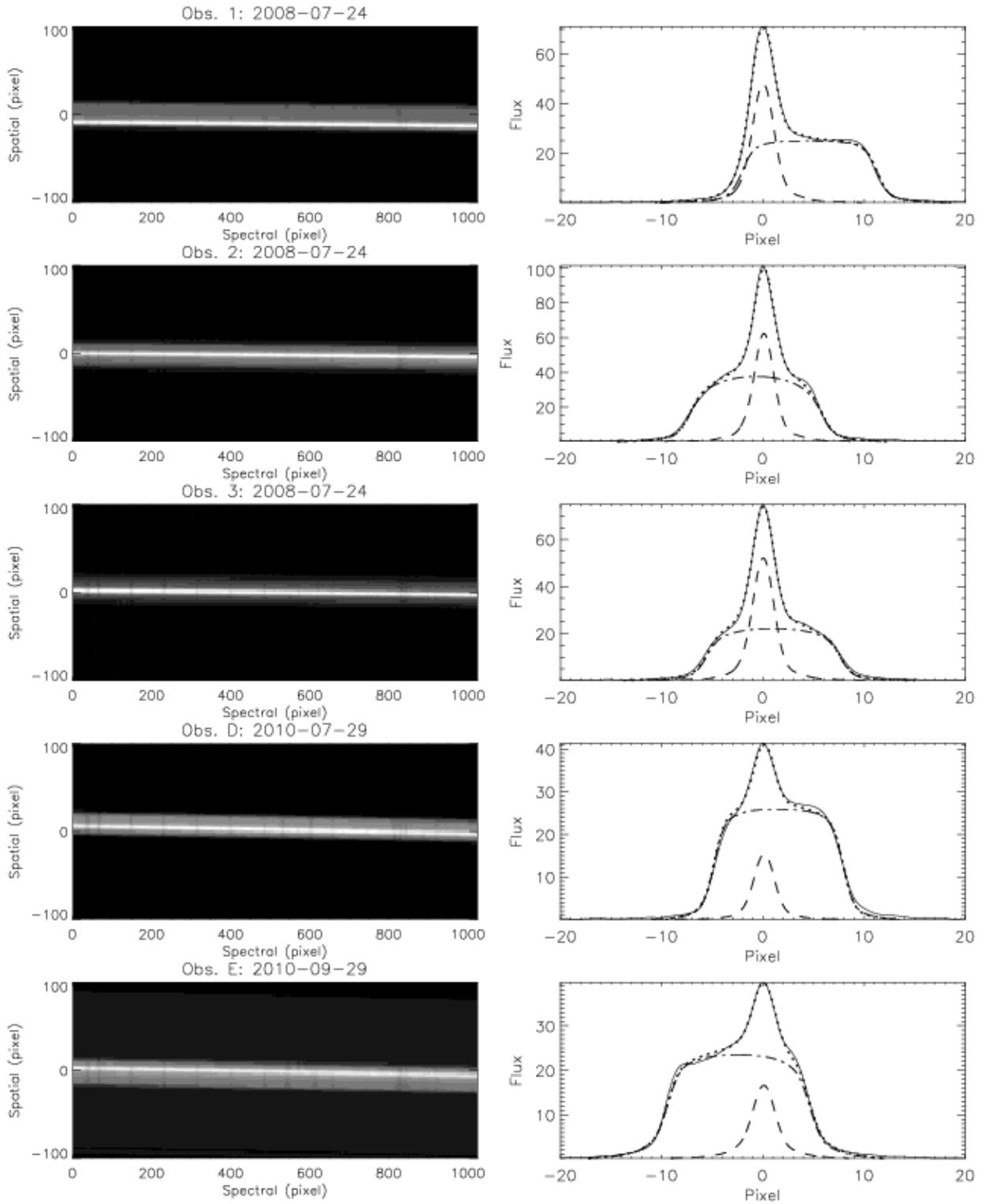

Fig. 1

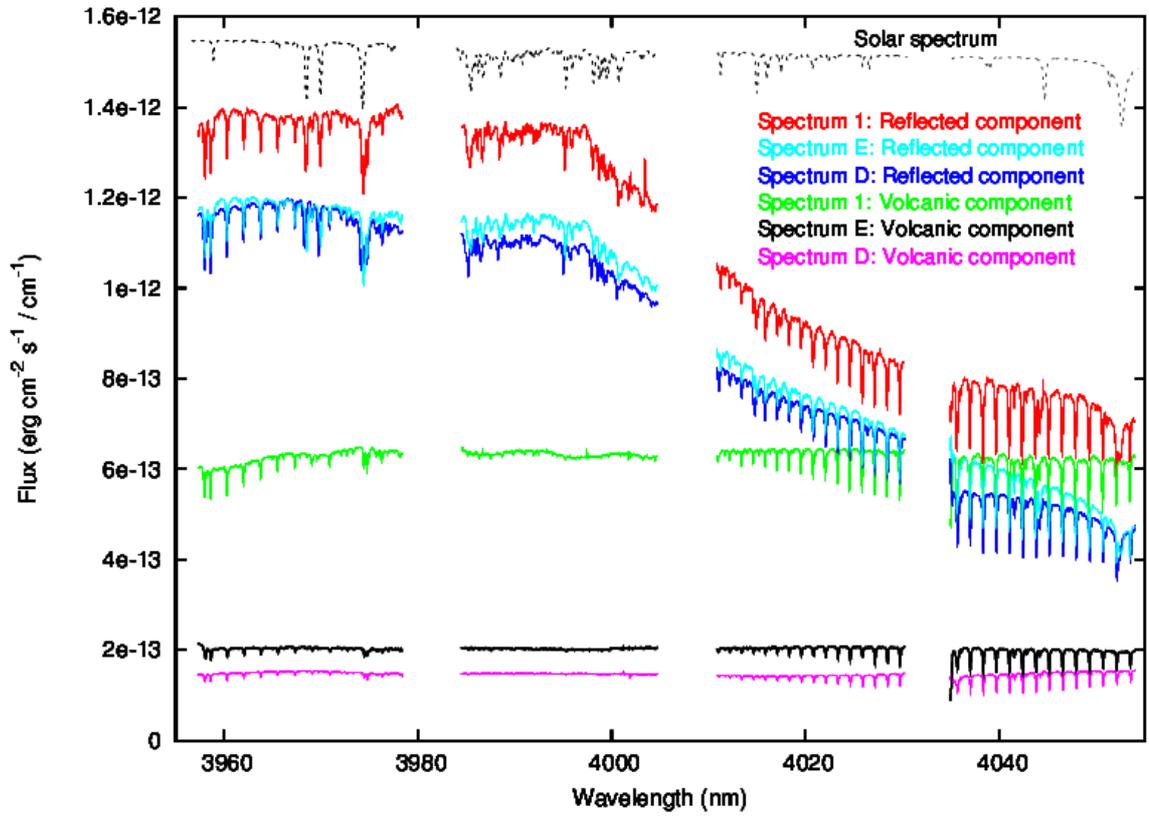

Fig.2

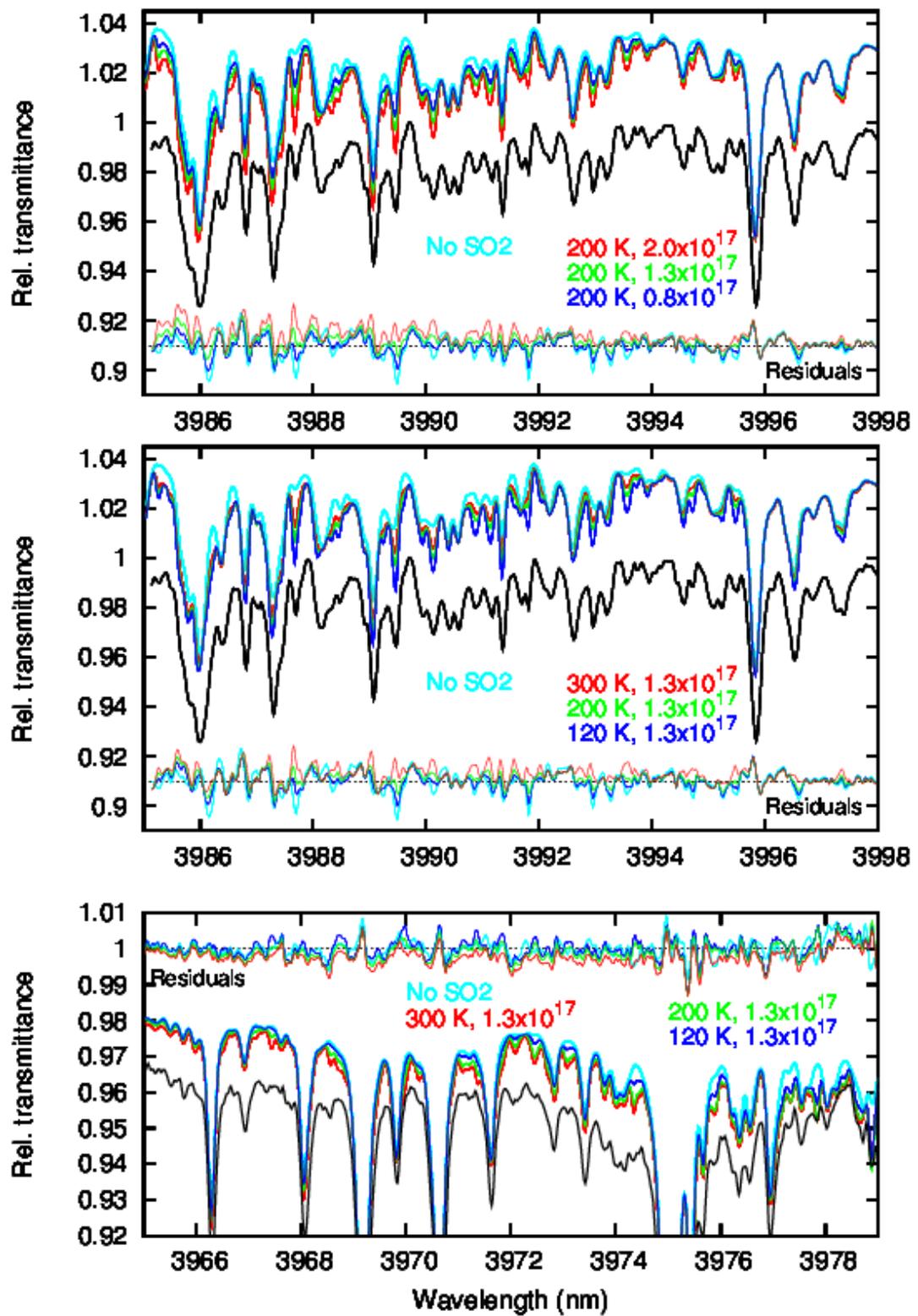

Fig. 3

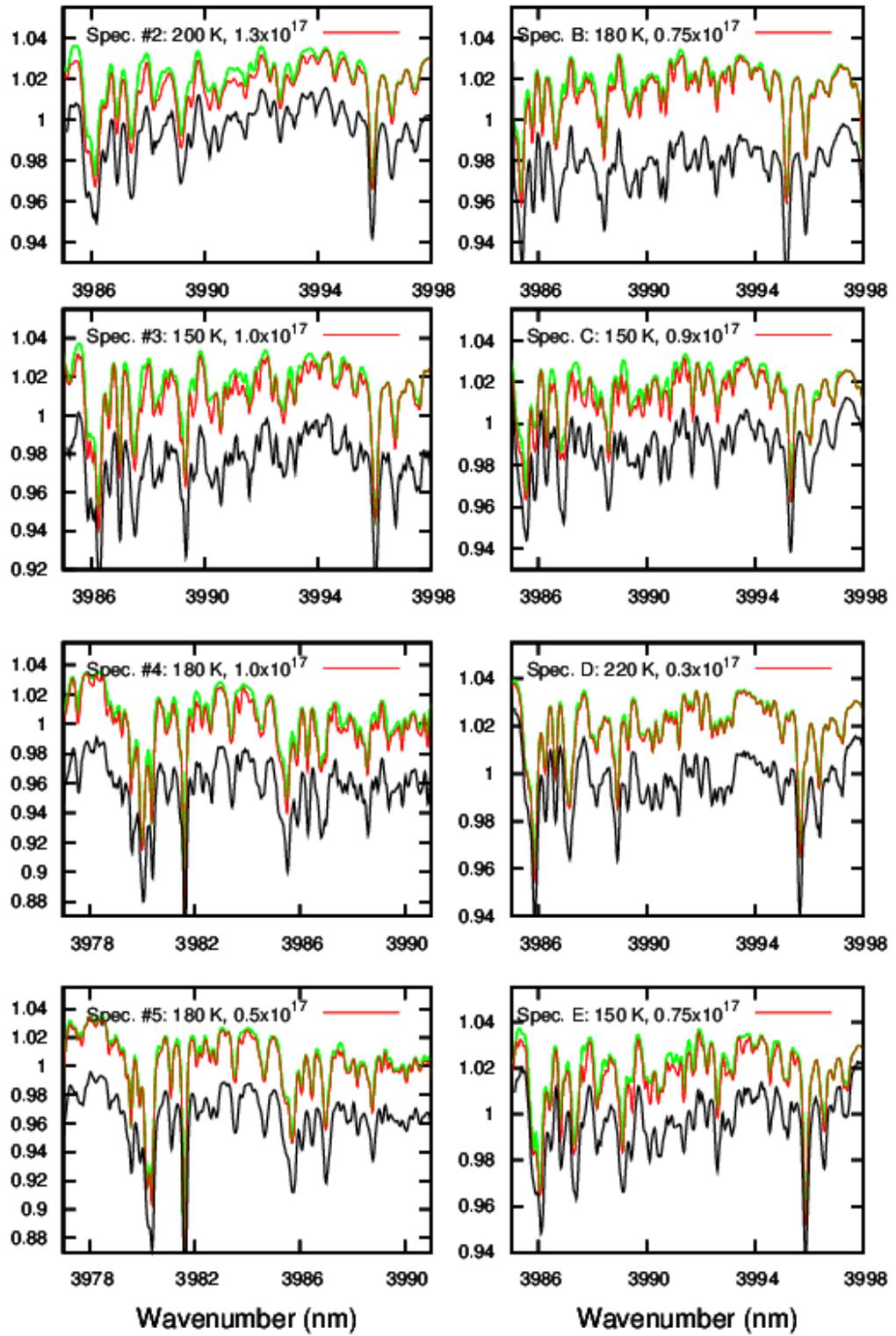

Fig. 4

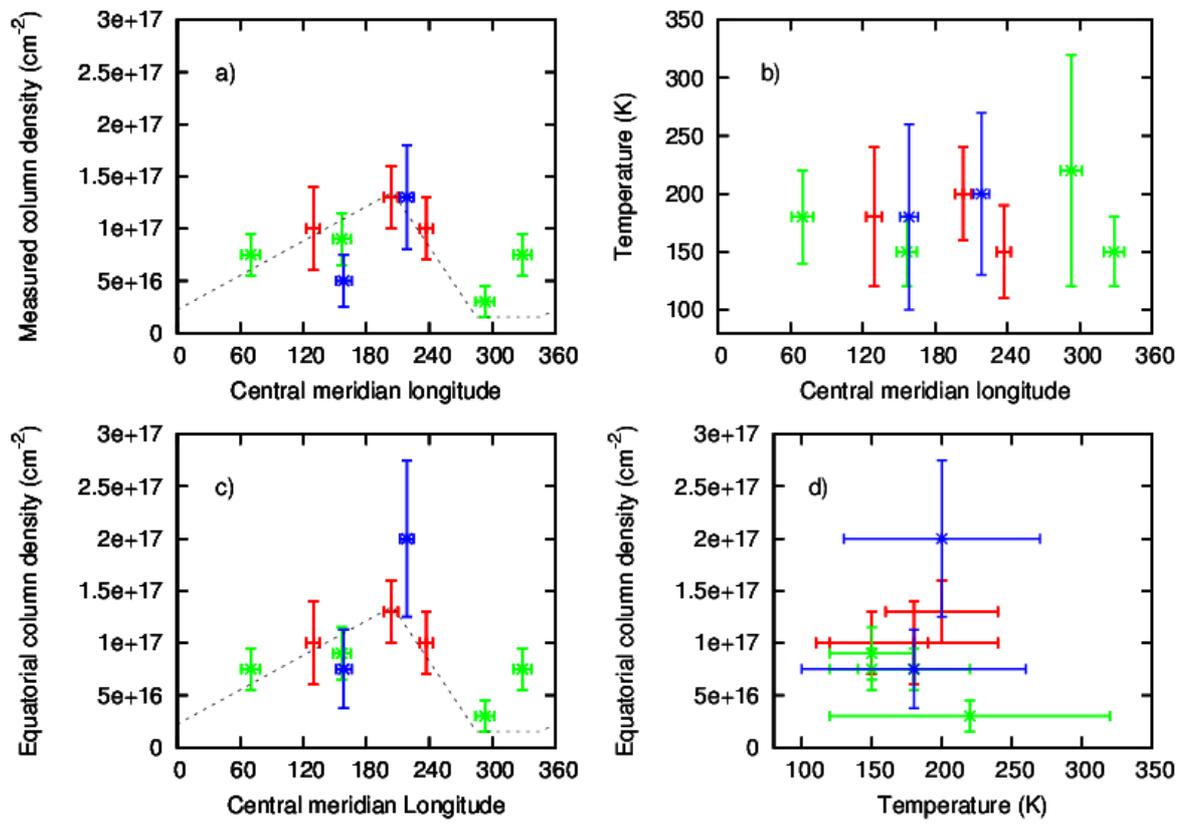

Fig.5

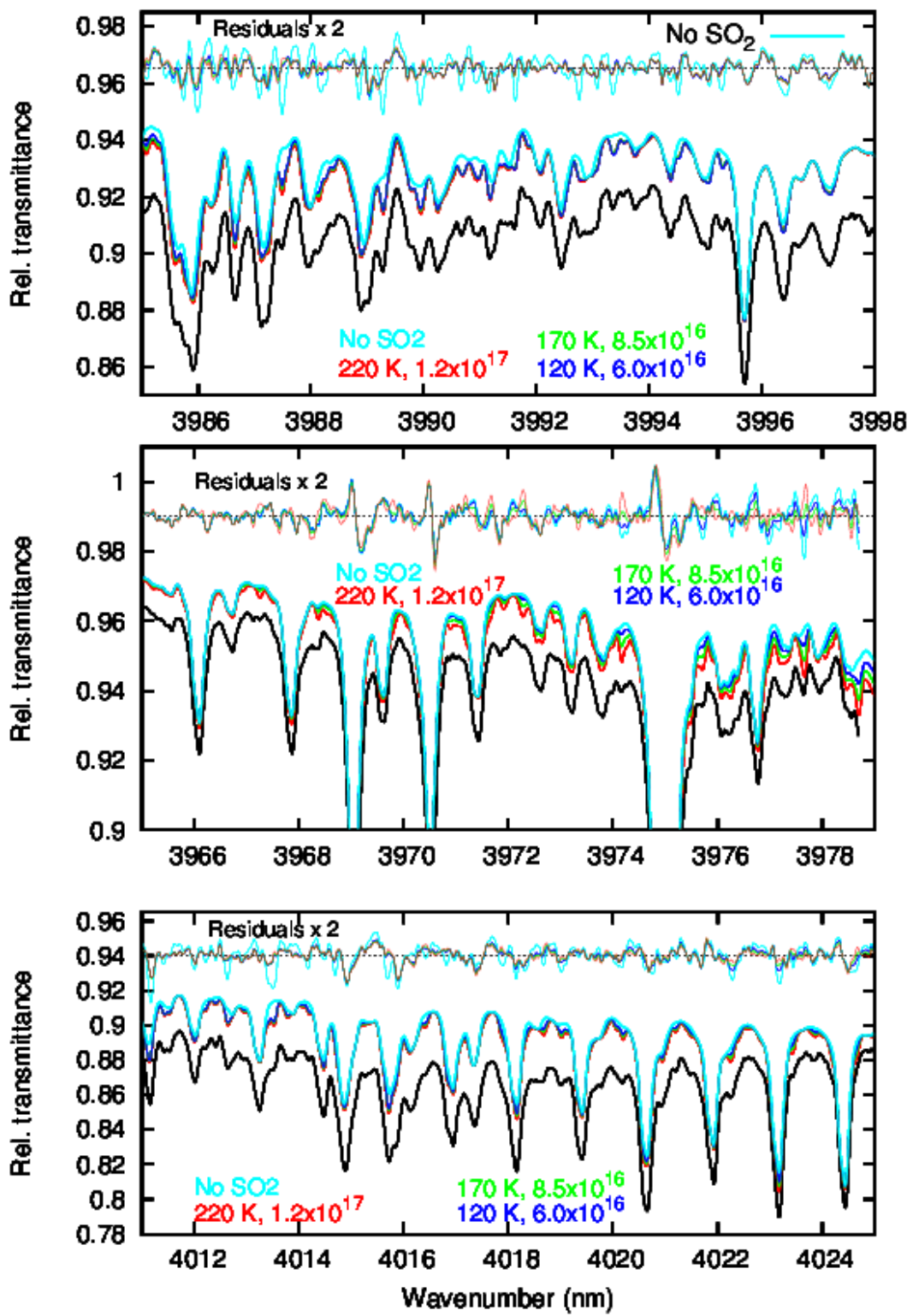

Fig. 6a

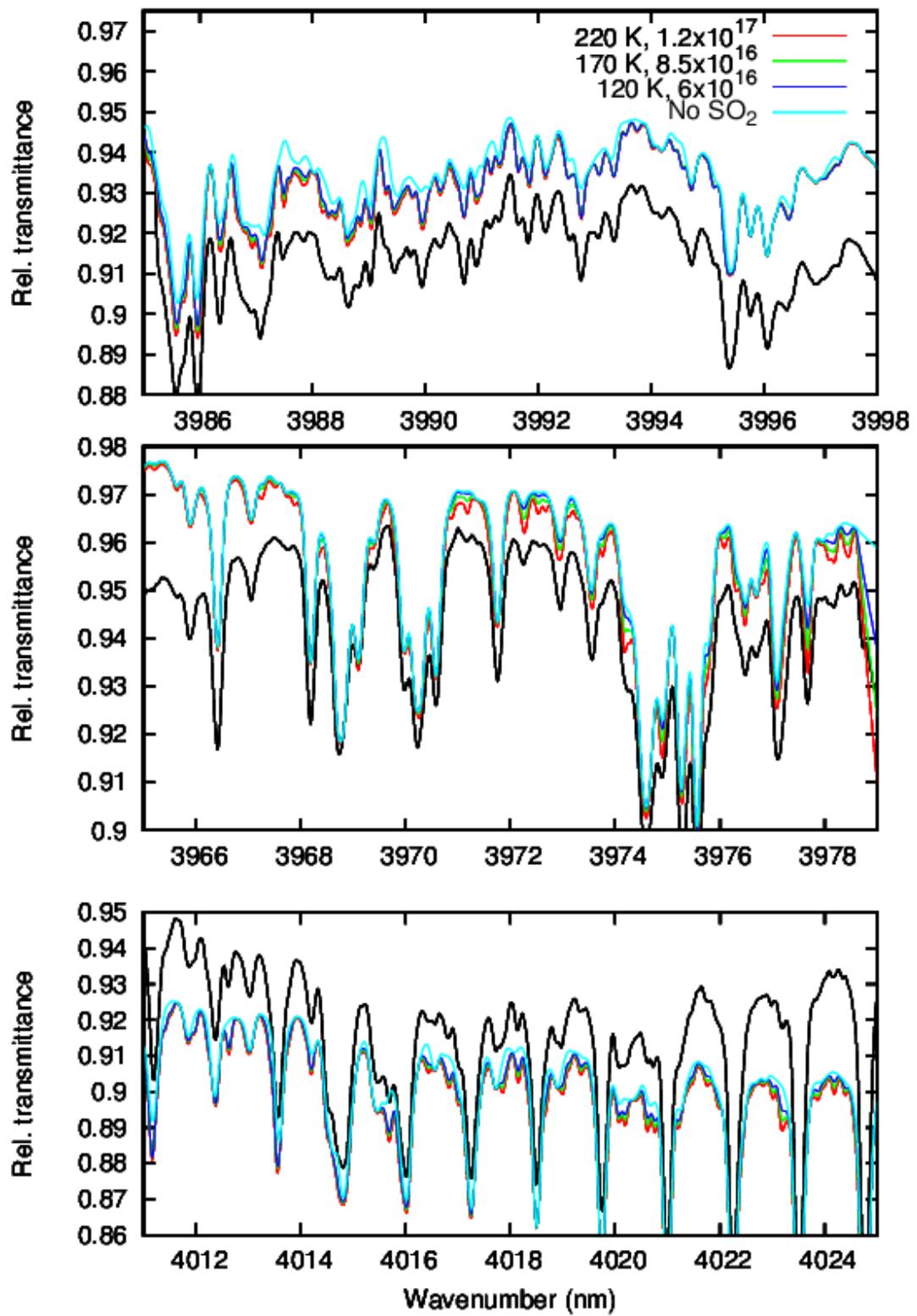

Fig 6b.

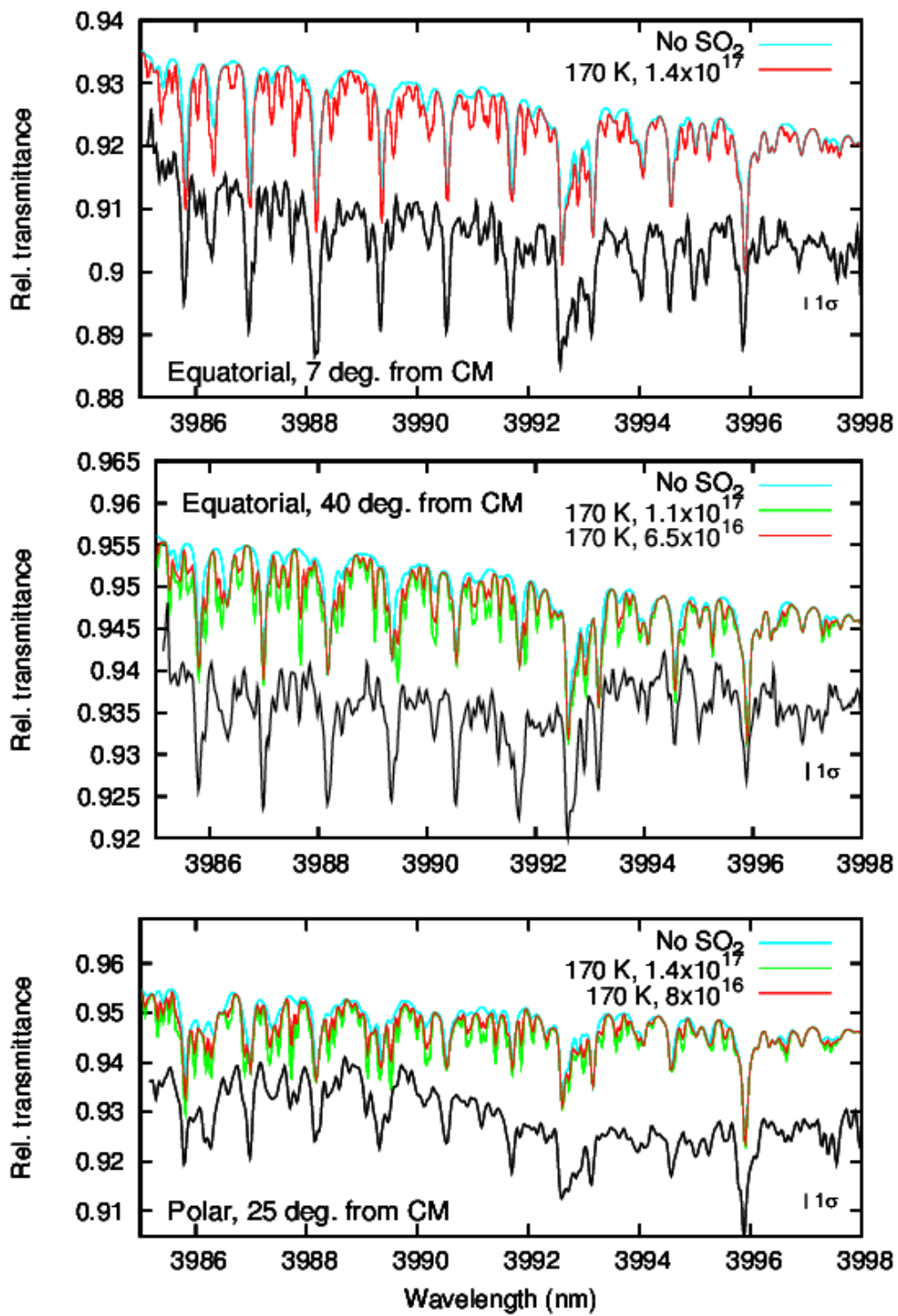

Fig. 7

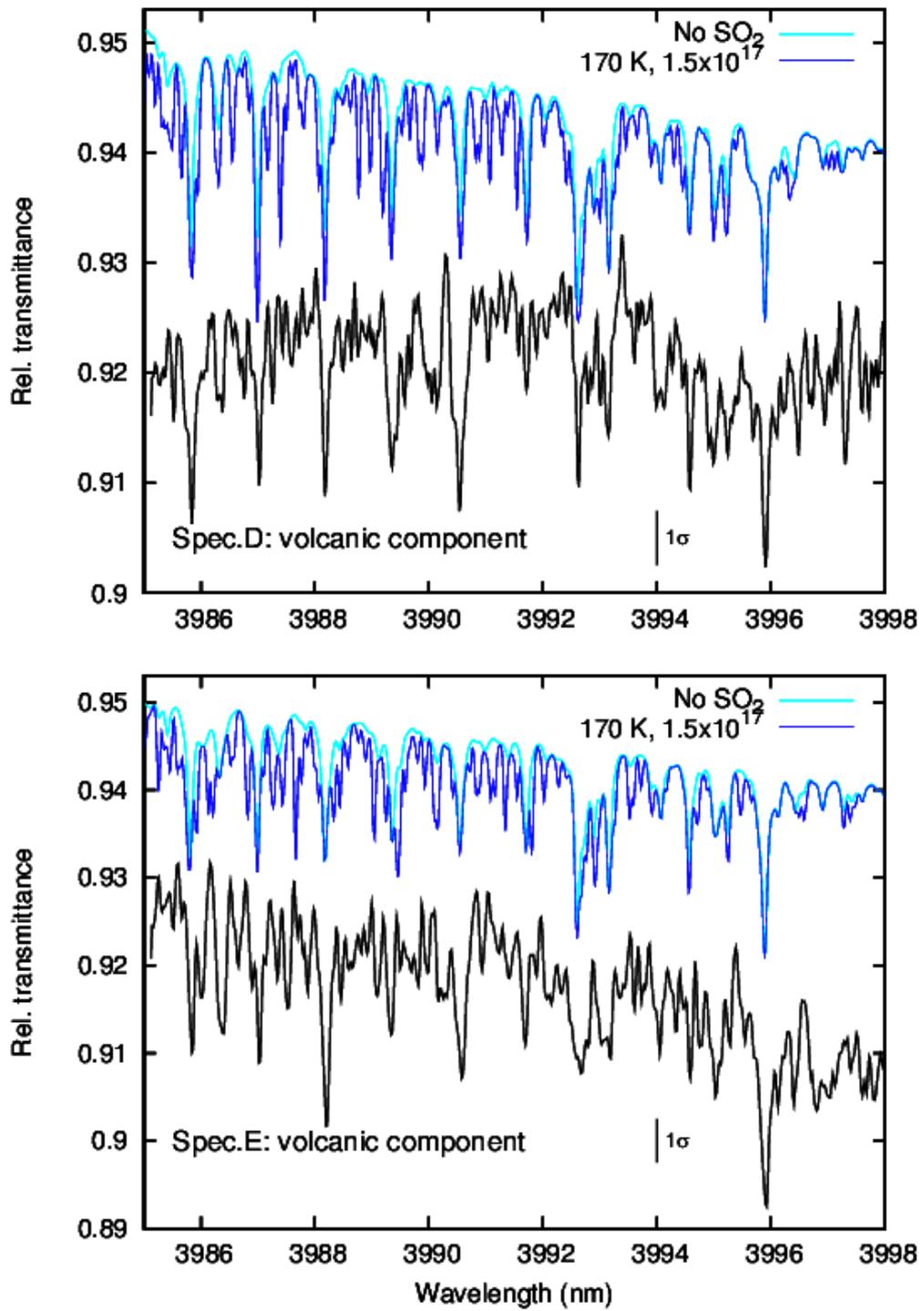

Fig. 8

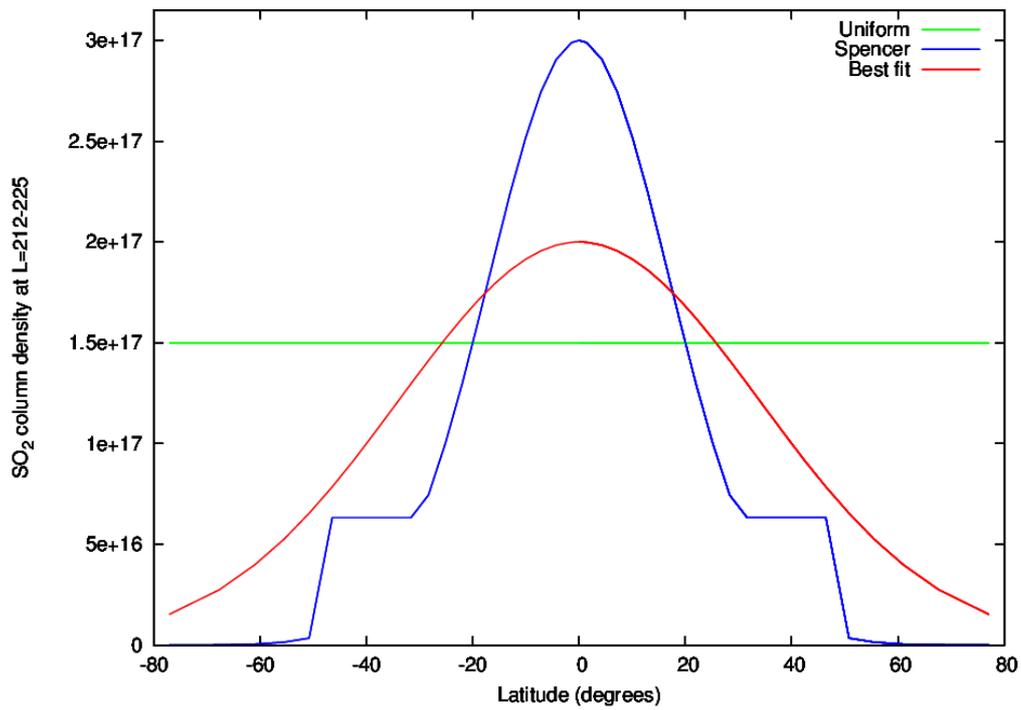

Fig. 9a

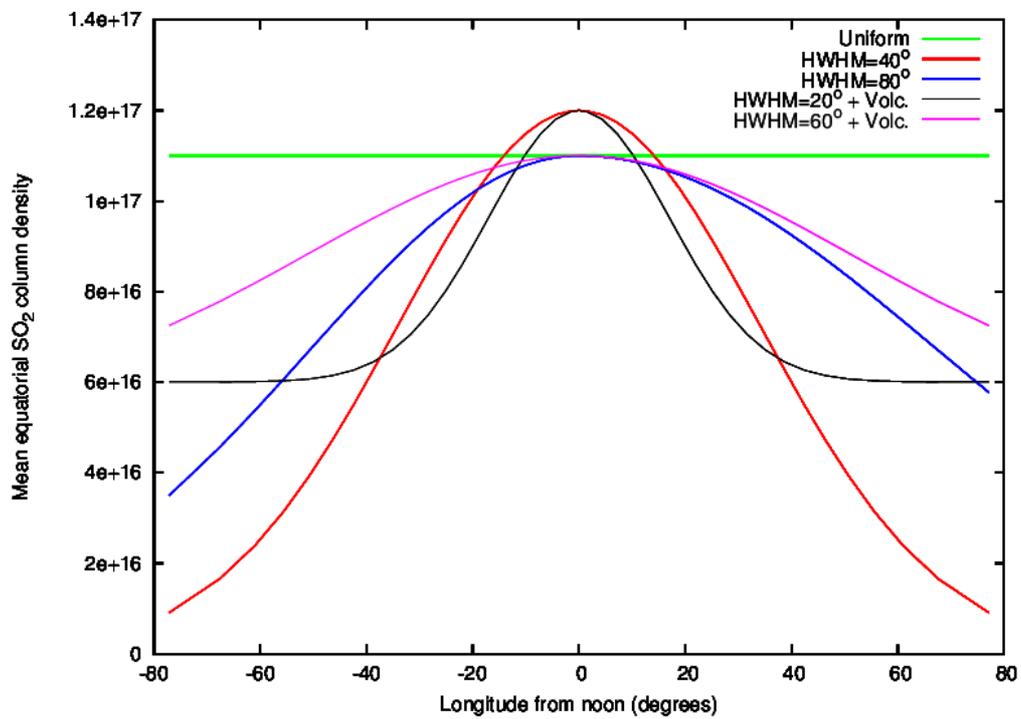

Fig. 9b

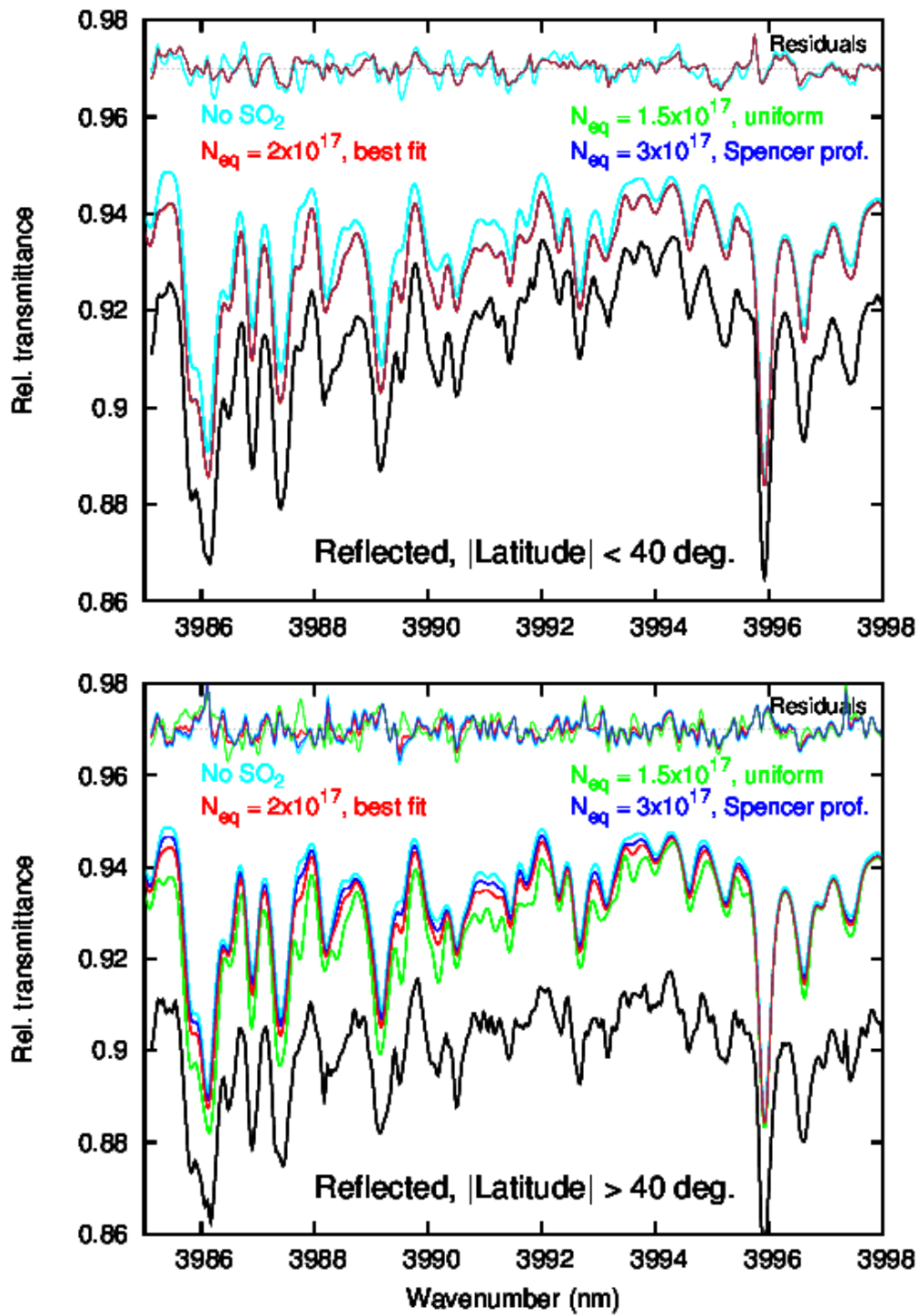

Fig. 10

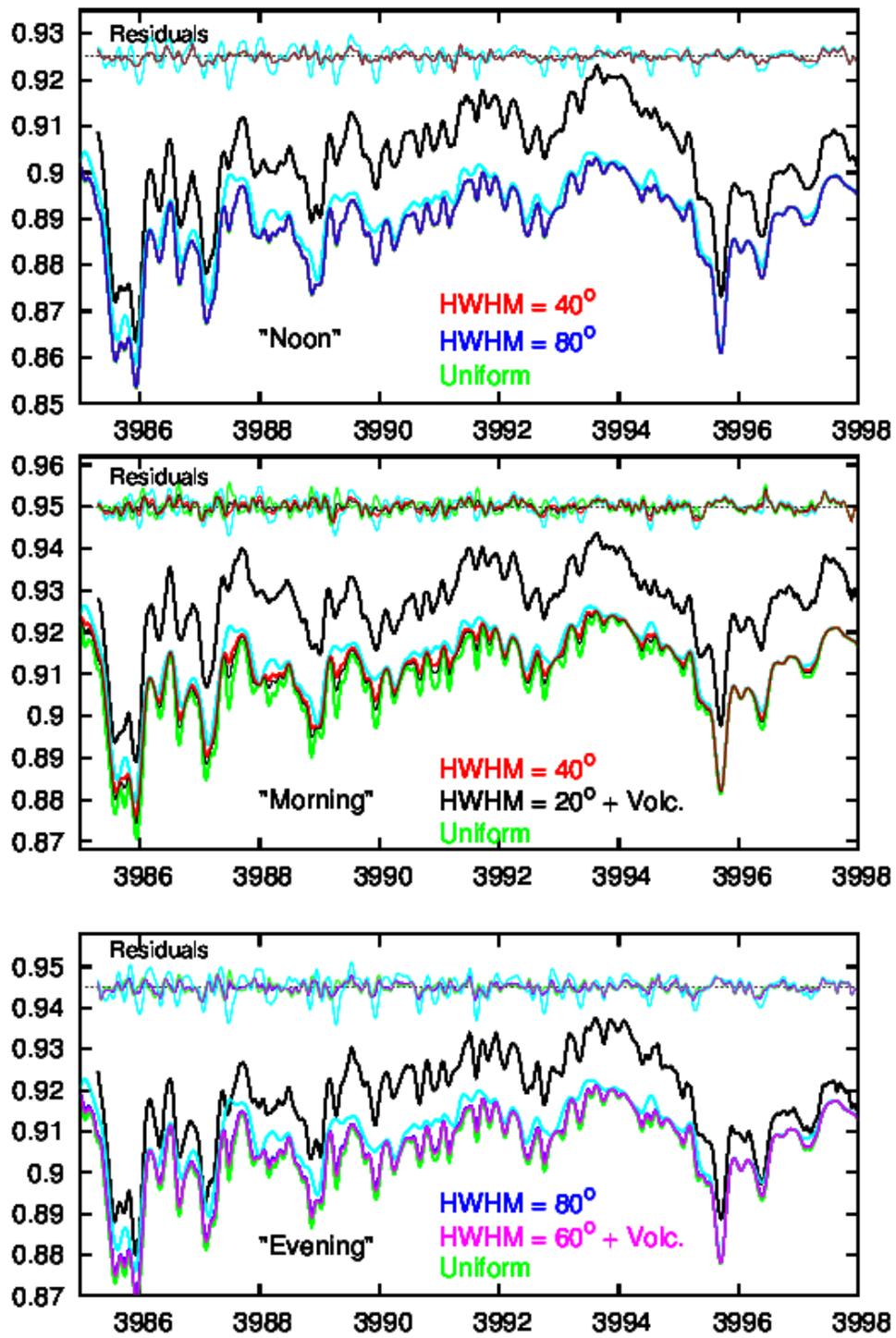

Fig. 11.

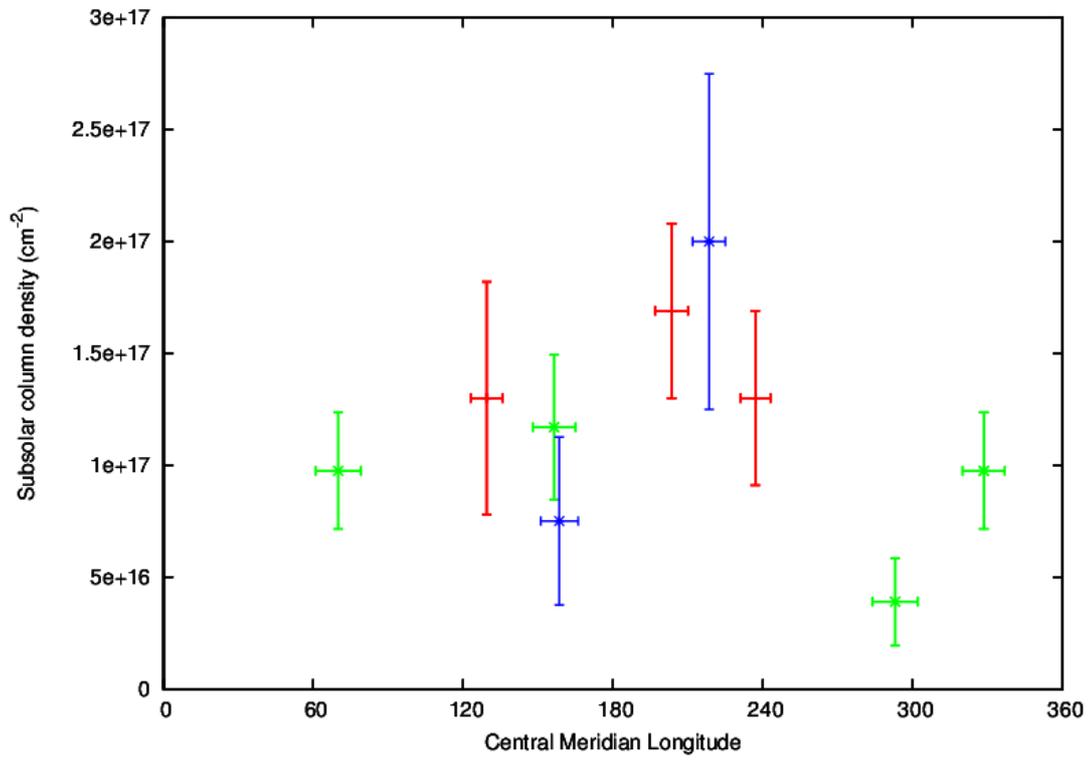

Fig. 12.